\documentclass[twocolumn,showpacs,showkeys,amsmath,amssymb]{revtex4}

\usepackage{graphicx}

\begin{document}

\title{Analytic Kerr black hole lensing for equatorial observers
in the strong deflection limit}

\author{V. Bozza$^{a,b,c}$, F. De Luca$^{b,c,d}$, G. Scarpetta$^{b,c,e}$, M. Sereno$^{c,f,g}$}

\affiliation{$^a$ Centro Studi e Ricerche ``Enrico Fermi'',
Compendio Viminale, I-00184, Rome, Italy.\\ $^b$ Dipartimento di
Fisica ``E.R. Caianiello'', Universit\`a di Salerno, via Allende,
I-84081 Baronissi (SA), Italy.\\
  $^c$  Istituto Nazionale di Fisica Nucleare, Sezione di
 Napoli. \\
 $^d$ Institut f\"{u}r Theoretische Physik der
           Universit\"{a}t Z\"{u}rich, CH-8057 Z\"{u}rich, Switzerland \\
 $^e$ International Institute for Advanced Scientific Studies, Vietri sul Mare (SA),
Italy \\
 $^{f}$ Dipartimento di Scienze Fisiche, Universit\`{a} degli Studi di
Napoli `Federico II', Via Cinthia, Monte S. Angelo, 80126 Napoli,
Italy. \\
 $^g$ Istituto Nazionale di Astrofisica -
Osservatorio Astronomico di Capodimonte, Salita Moiariello, 16,
80131 Napoli, Italy.}

\date{\today}

\begin{abstract}
In this paper we present an analytical treatment of gravitational
lensing by Kerr black holes in the limit of very large deflection
angles, restricting to observers in the equatorial plane. We
accomplish our objective starting from the Schwarzschild black
hole and adding corrections up to second order in the black hole
spin. This is sufficient to provide a full description of all
caustics and the inversion of lens mapping for sources near them.
On the basis of these formulae we argue that relativistic images
of Low Mass X-ray Binaries around Sgr A* are very likely to be
seen by future X-ray interferometry missions.
\end{abstract}



\maketitle

\section{Introduction}

General relativity predicts that light rays passing close to a
black hole suffer gravitational lensing, so that an observer
almost aligned with the line connecting a source and a black hole
sees two images of the original source. These images are due to
small deviations of photons that pass far enough from the black
hole to allow a weak field approximation of the metric tensor.
However, already Darwin in 1959 noticed that photons passing very
close to a black hole may suffer much larger deflections without
falling into the event horizon \cite{Dar}. In principle, an
observer situated on the line connecting the source and the black
hole, besides the two classical weak field images, would detect
two infinite series of images very close the black hole. These
images are produced by photons making one or more complete loops
around the black hole before re-emerging in the observer
direction. Of course, these relativistic images are largely
demagnified w.r.t the original source and for some time they just
remained a mathematical curiosity of General Relativity.

Nevertheless, things changed after the great progress of
interferometric techniques and the widely-accepted opinion that
the radio-source Sgr A* in the Galactic center actually hosts a
supermassive black hole of $3.61\times 10^5$ solar masses
\cite{Eis} (for a review see \cite{MelFal}). These facts motivated
Virbhadra \& Ellis to propose that this black hole may be an ideal
candidate for the generation of relativistic images of sources
eventually passing behind it \cite{VirEll}. In fact, the angular
radius of the shadow of Sgr A* is predicted to be 23 $\mu$as,
which is comparable to the best resolution achieved in the
millimeter band (18 $\mu$as \cite{Krich}). A complete imaging in
the sub-mm band was suggested in Ref. \cite{FMA}. Future space
missions in the infrared and in the X-rays may reach even higher
resolutions (for a complete discussion, see Ref. \cite{S1-S14}).

A new field of Gravitational lensing was definitively opened and
several authors proposed alternative methods to overcome the
evident difficulties of full general relativity calculations of
geodesics which typically result in cumbersome equations and heavy
numerical integrations \cite{Others,Perlick}. However, Darwin
himself proposed a surprisingly easy formula for the positions of
the relativistic images generated by a Schwarzschild black hole
\cite{Dar}. This formula and its consequences were later discussed
or re-discovered several times \cite{Atk,Lum,Oha,Nem} before
Virbhadra \& Ellis proposal. After that work, it was revived in
Ref. \cite{BCIS}, where it was called the strong field limit of
the deflection angle. It was then extended to Reissner-Nordstrom
black holes in Ref. \cite{ERT} and applied to microlensing by Sgr
A* by Petters \cite{Pet}. In this paper, as suggested by Perlick
\cite{Perlick}, we shall revise this terminology, referring to a
Strong Deflection Limit (SDL), since an infinite deflection angle
is not necessarily related to a large curvature. This can be
realized by considering a very large black hole. The minimum
distance reachable by a photon without being captured is of the
same order of the horizon radius. The Riemann invariant
$R_{\alpha\beta\gamma\delta}R^{\alpha\beta\gamma\delta}$,
evaluated in the inner region probed by the photon, scales as the
curvature at the horizon, i.e. $1/M^4$. Increasing the mass of the
black hole, the curvature felt by the photon becomes arbitrarily
low, even if its deflection may be large. So, it is more correct
to speak of a strong deflection limit without referring to the
curvature.

The power of the SDL expansion of the deflection angle became
evident when its universality was demonstrated in Ref.
\cite{Boz1}. Any class of spherically symmetric black holes leads
to the same SDL expansion; the coefficients of this expansion
depend on the specific class of the black hole, representing a
sort of identity card, from which all the parameters of the black
hole can be extracted. By observing the relativistic images of a
gravitational lensing event induced by a black hole, it is
possible, in principle, to deduce all its parameters and
properties. Since this could also provide the key to discriminate
between General Relativity and some extended theories of
gravitation, this method has been applied to several interesting
classes of black holes coming from string theory, braneworlds and
wormholes \cite{StrBra}. Some limitations were removed in Refs.
\cite{Retro,BozMan}, while time delay analysis was performed in
Ref. \cite{TimDel}.

As regards spinning black holes, the story is more complicated.
Almost forty years have passed since Carter reduced the geodesics
equations in Kerr spacetime to first order equations depending on
four constants of motion \cite{Car}. This fundamental achievement
allowed a complete study and classification of all possible
trajectories of massive particles and photons moving around
spinning black holes \cite{Cha}. In order to visualize and study
these geodesics, a very large amount of numerical methods has been
developed through years. In the context of gravitational lensing,
these methods have been used to describe the light curve of a star
orbiting a black hole \cite{CunBar} and the apparent shape of the
accretion disk \cite{Lum,Accret}. Rauch \& Blandford have proved
the formation of extended 4-cusped caustics numerically
\cite{RauBla}.

Extending the SDL methodology to axially symmetric black holes is
not immediate and the simplicity of the approach may be easily
lost. In Ref. \cite{BozEq} the SDL formula was recovered for light
rays lying close to the equatorial plane of a Kerr black hole, but
the coefficients of the formula had to be calculated numerically
as functions of the black hole spin. Vazquez \& Esteban solved the
lens equation far from the equatorial plane for some particular
cases \cite{VazEst}, but a complete analytical treatment of Kerr
lensing is still missing.

In this paper we make a considerable step toward this objective,
focusing on observers lying on the equatorial plane and solving
the general lens equation for small values of the black hole spin.
Perturbative methods allow us to use the Schwarzschild SDL formula
as starting point to describe the deflection of light rays looping
around a Kerr black hole in a completely analytical way. Our
treatment leads to an amazingly simple description of all
relativistic caustics and to the immediate inversion of lens
mapping for sources near caustics. The limitation to the
equatorial observer is motivated by the fact that the most
important candidate black hole, Sgr A*, is likely to have a spin
axis perpendicular to the galactic plane, where the solar system
lies, in a first approximation. It is natural, then, to take
advantage of this configuration and deal with considerably
simplified equations.

Our paper is structured as follows. Sect. 2 recalls the main
results of Kerr geodesics. Sect. 3 explains how the SDL is
introduced in Kerr gravitational lensing and illustrates the
strategy we use to solve the geodesics equations. Sect. 4 contains
the derivation of the caustics order by order. In Sect. 5 we
analyze the lens map close to the relativistic caustics, finding
the additional images and their magnification. Sect. 6 considers
the perspectives for observations in the light of what we have
found, focusing on the study of relativistic images around Sgr A*.
Sect. 7 summarizes the main results of the paper. Two appendices
complement the calculations explained in Sect. 3 with more
details.

\section{Kerr geodesics}

In Boyer-Lindquist coordinates \cite{BoyLin} $x^\mu \equiv
(t,x,\vartheta,\phi)$, the Kerr metric reads

\begin{eqnarray}
& ds^2=&\frac{\Delta-a^2 \sin^2 \vartheta}{\rho^2}d
t^2-\frac{\rho^2}{\Delta} dx^2-\rho^2 d\vartheta^2 \nonumber \\
&& - \frac{ \left(x^2+a^2 \right)^2 - a^2\Delta \sin^2 \vartheta
}{\rho^2} \sin^2 \vartheta d\phi^2 \nonumber \\&&+\frac{2ax
\sin^2\vartheta}{\rho^2} dt d\phi \\%
& \Delta=&x^2-x+a^2 \\%
& \rho^2=& x^2+a^2 \cos^2\vartheta
\end{eqnarray}
where $a$ is the specific angular momentum of the black hole. All
distances are measured in Schwarzschild radii ($2MG/c^2=1$).
$\vartheta$ and $\phi$ represent the polar and azimuthal angles
respectively, while $x$ is the radial coordinate. The event
horizon is a spherical surface of radius
$x_H=(1+\sqrt{1-4a^2})/2$. In our notations, $a$ runs from 0
(Schwarzschild black hole) to $1/2$ (extremal Kerr black hole).

Carter showed that the Kerr geodesics can be resolved in terms of
first integrals of motion \cite{Car}. The final expressions for
lightlike geodesics take the following form (following Ref.
\cite{Cha})
\begin{eqnarray}
&& \pm \int \frac{dx}{\sqrt{R}}=\pm \int \frac{d
\vartheta}{\sqrt{\Theta}} \label{Geod1}\\
& \phi_f-\phi_0 =& a \int\frac{x^{2}+a^{2}-a J}{\Delta \sqrt{R}}
dx-a \int \frac{dx}{\sqrt{R}} \nonumber  \\
&& + J \int \frac{\csc^2\vartheta}{\sqrt{\Theta}} d \vartheta,
\label{Geod2}
\end{eqnarray}
where
\begin{eqnarray}
&\Theta=&Q+a^2 \cos^2\vartheta-J^2 \cot^2\vartheta \\
&R=&x^4+(a^2-J^2-Q)x^2+(Q+(J-a)^2) x \nonumber \\ &&-a^2 Q.
\label{R}
\end{eqnarray}
In these expressions, $J$ and $Q$ are two constants of motion
that, along with the initial condition $\phi_0$, completely
identify the geodesic. The double signs in front of the integrals
in Eq. (\ref{Geod1}) remind that the integrals must be performed
piecewise, between two consecutive values of $x$ and $\vartheta$
that annihilate the denominators $R$ and $\Theta$ (inversion
points). Then the sign of each piece between two inversion points
is chosen in such a way that all of them sum up with the same
sign, giving a final positive contribution.

In the context of gravitational lensing, we are interested to
those photons that come from an infinite distance, approach the
black hole reaching a minimum distance $x_0$ and then escape back
to infinity. This selects trajectories characterized by $Q \geq
0$. Moreover, since the roots of $R$ represent inversion points in
the radial motion, we have to impose that $R$ has one
non-degenerate positive root. This amounts to require that
$R(x_0)=0$, $R'(x_0) \neq 0$. The limiting situation, when $x_0$
becomes a degenerate root, is obtained when the equations
$R(x_m)=0$, $R'(x_m) = 0$ are simultaneously fulfilled at some
point $x_m$. Solving these equations w.r.t. $J$ and $Q$ we get
\begin{eqnarray}
&&J_{m}=\frac{x_{m}^{2}(-3+2 x_{m})+a^{2}(1+2 x_{m})}{a(1-2
x_{m})} \label{Jm}\\
&& Q_{m}=\frac{x_{m}^{3}\left[ 2
a^2-x_{m}(x_{m}-3/2)^2\right]}{a^{2}(x_{m}-1/2)^2}. \label{Qm}
\end{eqnarray}
Given a value of $x_m$, the quantities $J_m$ and $Q_m$ represent
the values of $J$ and $Q$ that characterize those trajectories
that bring a photon down to the distance $x_m$ in an infinite
time. Asymptotically the photon keeps orbiting forever at a fixed
distance $x_m$ from the black hole. However, this orbit is
unstable and small perturbations make the photon fall into the
black hole or deviate it back to infinity. In Schwarzschild black
hole, the radius $x_m$ of the unstable photon orbit is fixed to
$3/2$ in Schwarzschild units (the sphere of radius $x_m$ is then
called photon sphere). In the case of Kerr black holes, the radius
of the orbit depends on the initial orientation of the photon
trajectory. In practice, $x_m$ may vary between two limiting
values $x_{m+}$, $x_{m-}$, which respectively represent the radius
of the orbit described by a photon co-rotating with the black hole
and the radius of the orbit of a counter-rotating photon in the
equatorial plane. All intermediate values correspond to photons
whose orbits are not equatorial and do not lie on a single plane.
In order to find these limiting values, we have to impose $Q_m =
0$. So, $x_{m+}$ and $x_{m-}$ are found as the two largest roots
of this equation. This is a third degree equation whose solution
can be found exactly. However, since the successive calculations
would soon become too cumbersome, we will already consider their
expansions in powers of $a$. To describe second order effects in
the lens equation, it is necessary to take terms up to the third
order:
\begin{equation}
x_{m\pm}=\frac{3}{2}\mp \frac{2}{\sqrt{3}} a -\frac{4}{9} a^2 \mp
\frac{20}{27\sqrt{3}}a^3+ O(a^4). \label{xmpm}
\end{equation}

We see that in the limit $a \rightarrow 0$, the two limiting
values converge to the Schwarzschild photon sphere $x_m=3/2$. When
$a$ is different from zero, $x_{m+}$ and $x_{m-}$ are distinct.
The specific value of $x_m$ in the interval $[x_{m+},x_{m-}]$
uniquely fixes the amplitude of the oscillations on the equatorial
plane performed by the photon along its orbit. In consideration of
this fact we introduce a more convenient parametrization,
replacing $a$ by $a \xi$:
\begin{equation}
x_{m}(\xi)=\frac{3}{2}-\frac{2}{\sqrt{3}} a \xi -\frac{4}{9} a^2
\xi^2 - \frac{20}{27\sqrt{3}}a^3\xi^3. \label{xmsv}
\end{equation}
Varying the parameter $\xi$ in the range $[-1,1]$ we obtain all
possible values of $x_m$ in the range $[x_{m+},x_{m-}]$,
corresponding to orbits with different amplitude of the
oscillations on the equatorial plane (a different parameterization
with similar properties was used in Ref. \cite{Zak}). We shall see
that all quantities assume very simple expressions in terms of
this parameter $\xi$. Now, using this parametrization in Eqs.
(\ref{Jm})-(\ref{Qm}), we can expand $J_m$ and $Q_m$ to second
order in $a$ and read them as functions of $\xi$:
\begin{eqnarray}
&J_m(\xi)=& \frac{3 \sqrt{3}}{2} \xi-2 a-\frac{a^{2} \xi
(2-\xi^{2})}{\sqrt{3}} + O(a^3), \label{Jma2} \\ &Q_{m}(\xi)=&
\frac{27}{4} (1-\xi^2)+3 a^{2} \xi^{2} (1-\xi^2)+ O(a^3).
\label{Qma2}
\end{eqnarray}
Notice that the presence of $a$ in the denominators of Eqs.
(\ref{Jm})-(\ref{Qm}) allows $\xi$ to appear at zero order
already. That is why we needed a third order expansion for $x_m$.
So, even in the Schwarzschild limit, $\xi$ can be used to
parametrize the photon sphere in the $(J,Q)$ plane.

\begin{figure}
\resizebox{\hsize}{!}{\includegraphics{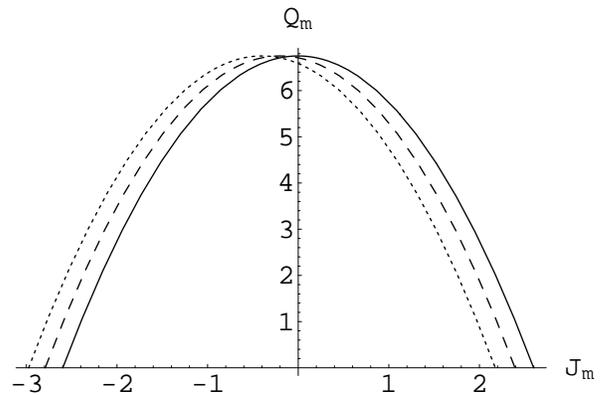}} \caption{ The
limiting values $J_m$ and $Q_m$ for the constants of motion $J$
and $Q$ corresponding to trajectories reaching the unstable orbit
around the black hole asymptotically. The solid line is for $a=0$,
the dashed line is for $a=0.1$ and the dotted line is for $a=0.2$.
 }
 \label{Fig JmQm}
\end{figure}

In Fig. \ref{Fig JmQm} we plot the locus of points $(J_m,Q_m)$
when we vary $\xi$ in the range $[-1,1]$, for different values of
$a$. We recall that purely prograde photons travelling on the
equatorial plane are characterized by $Q=0$ and positive $J$,
while retrograde photons have negative $J$. Photons with $J=0$ and
$Q>0$ run on polar trajectories. Any geodesics characterized by
$J$ and $Q$ outside this locus (with $Q \geq 0$), correspond to
acceptable gravitational lensing trajectories. All photons with
$J$ and $Q$ inside this locus are destined to fall inside the
black hole.

There is an immediate connection between these constants and the
point in the sky where the observer detects the photon. Throughout
the paper, we consider a static observer at a radial
Boyer-Lindquist coordinate $D_{OL}$, lying on the equatorial
plane. This restriction will keep all equations simple enough to
be solved, while ensuring an exhaustive description of the
expected phenomenology for Sgr A*. This definition has no
ambiguity from a mathematical point of view, but needs to be
linked to the astrophysical notion of distance from the solar
system to Sgr A*. The current measurements of the distance to the
Galactic center are typically based on the dynamical investigation
of the stars orbiting around Sgr A*. The orbital fits are done in
the context of Newtonian gravity. As a consequence, the current
estimate of the distance to the Galactic center, which amounts to
about 8 kpc \cite{Reid,Eis}, assumes a flat background geometry.
This flat distance makes sense as long as all scales are much
larger than the Schwarzschild radius of the central black hole,
which is
\begin{equation}
R_\mathrm{Sch}=\frac{2G M}{c^2} = 1.1 {\times} 10^{10} ~
\mathrm{m}
\end{equation}
for $M=3.61\times 10^6$ M$_\odot$ \citep{Eis}. Now it is evident
that in the limit of large distances, one can simply translate any
flat distance from the black hole, as calculated by Newtonian
physics, into Boyer-Lindquist coordinates in the Kerr geometry. In
fact, far from the black hole, in the asymptotic region, the
Boyer-Lindquist coordinate coincide with the euclidean spherical
coordinates centered on the black hole. The ambiguity in this
identification is of the order of $R_{Sch}/x$ ($x$ being the
distance from the black hole) and becomes relevant only close to
the event horizon, where Newtonian physics loses any meaning. So,
we can safely assume $D_{OL}=8$ kpc, when speaking about Sgr A* in
any calculations.

\begin{figure}
\resizebox{\hsize}{!}{\includegraphics{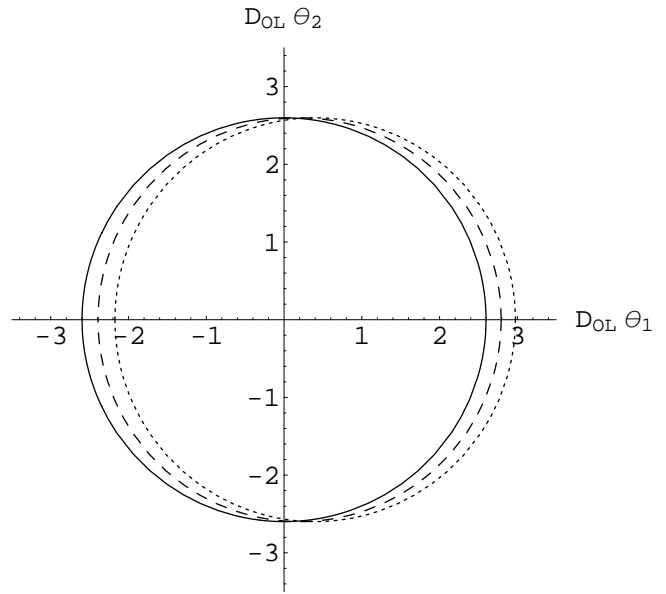}} \caption{ The
shadow shape in the observer sky. The solid line is for $a=0$, the
dashed line is for $a=0.1$ and the dotted line is for $a=0.2$.
 }
 \label{Fig Thm}
\end{figure}

Then, considering only observers in the asymptotic region ($D_{OL}
\gg 1$), where the geometry is close to be Minkowskian, it is
possible to define angular coordinates $(\theta_1,\theta_2)$ in
the observer sky. We will put the black hole in $(0,0)$, and let
$\theta_1$ run parallel to the equatorial plane of the black hole
while $\theta_2$ will run on the perpendicular axis (see Fig.
\ref{Fig Thm}). As $D_{OL} \gg 1$, $\theta_1$ and $\theta_2$ will
always be assumed to be very small. As shown in Ref. \cite{Cha},
photons reaching the observer from the generic point
$(\theta_1,\theta_2)$ are characterized by the constants
\begin{eqnarray}
&& J=-\theta_{1} D_{OL} \label{JTh1} \\
&& Q=\theta_{2}^2 D_{OL}^2 \label{QTh2}
\end{eqnarray}
w.r.t. to the black hole. We have taken the spin axis of the black
hole parallel to the $\theta_2$ direction and we have considered a
photon moving toward the observer. Then, it is immediate to pass
from $(J,Q)$ to the corresponding coordinates in the observer sky
$(\theta_1,\theta_2)$ and viceversa, apart from an ambiguity of
sign in $\theta_2$.

We can use these formulae to translate the locus $(J_m,Q_m)$ into
a new one $(\theta_{1,m},\theta_{2,m})$ in the plane
$(\theta_1,\theta_2)$. This is given by
\begin{eqnarray}
& \!\!\!\!\!\!\!\!\!\!\!\!\!\!\!\! D_{OL} \theta_{1,m}(\xi)=&
-\frac{3 \sqrt{3} \, \xi}{2}+2 a+\frac{a^{2} \xi
(2-\xi^{2})}{\sqrt{3}} +O(a^3)\label{Th1m}\\ & \!\!\!\!\!\!\!\!
\!\!\!\!\!\!\!\!D_{OL} \theta_{2,m}(\xi) =& \pm \left( \frac{3
\sqrt{3}}{2}+\frac{a^2 \xi^2}{\sqrt{3}} \right)
\sqrt{1-\xi^2}+O(a^3). \label{Th2m}
\end{eqnarray}
and is called the shadow of the black hole.

Fig. \ref{Fig Thm} shows the shape of the shadow in the observer
sky for different values of $a$. From what we have said before,
all photons deflected by the black hole must reach the observer
from directions $(\theta_1,\theta_2)$ outside the shadow. Photons
reaching the observer from the inside of the shadow cannot come
from gravitational deflection but must have been generated by
sources in front of the black hole. So, if we had enough
resolution to fully image a black hole, we would see a black
shadow with the shape described by Eqs. (\ref{Th1m})
and(\ref{Th2m}), bordered by a luminous ring due to gravitational
lensing of all sources around the black hole \cite{FMA}.

In Fig. \ref{Fig Thm} we see that the Schwarzschild shadow is
circular. Increasing the black hole spin $a$, the shadow is
slightly distorted and gets displaced to the right, meaning that
prograde photons (coming from the left side as seen from the
observer) are allowed to get closer to the black hole, while
retrograde photons (coming from the right side) must keep farther.

\begin{figure}
\resizebox{\hsize}{!}{\includegraphics{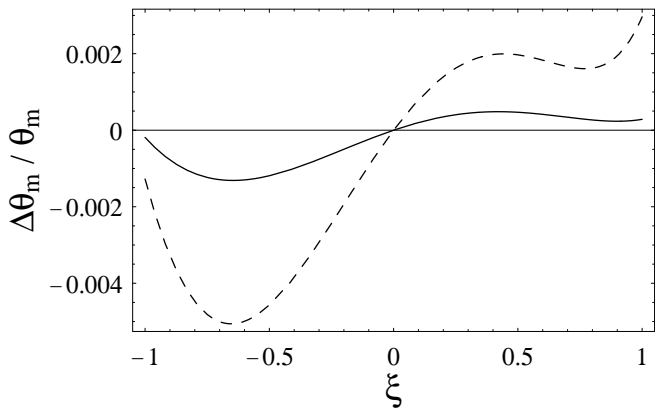}} \caption{
The relative variation in the radial distance of the black hole
shadow shape between the exact solution and the second order
approximation as a function of the azimuthal variable $\xi$. The
solid line is for $a =0.1$, the dashed line is for $a=0.2$.
 }
 \label{FigDeltaThm}
\end{figure}

Here, for later convenience, we are introducing and making use of
expressions expanded to the second order in $a$. However, the
exact expression for the shadow can be easily derived, combining
Eqs. (\ref{Jm})-(\ref{Qm}) with Eqs. (\ref{JTh1})-(\ref{QTh2}).
Comparing the exact shadow to its second order approximation, we
find that the latter works surprisingly well up to very high
values of the black hole spin. In Fig.~\ref{FigDeltaThm}, we plot
the relative error in the radial angular distance of the apparent
shadow $\theta_m$ in the approximate solution w.r.t the exact one
as a function of the variable $\xi$, which follows the azimuthal
angle in the observer's sky. Up to $a \lesssim 0.28$, relative
variations are well under 1\%, at $a=0.4$ we find an error of 2\%,
while in the extremal case $a=0.5$ the error only reaches 5\%.
Such errors must be compared to the displacement of the
relativistic rings from the shadow, see Sect. IV, and turn out to
be negligible for the higher order critical curves up to large
values of the spin.

\section{Kerr lensing in the Strong Deflection Limit}

It is useful to introduce the following parametrization:
\begin{equation}
\left\{ \begin{array}{l}
 \theta_1(\epsilon,\xi)=\theta_{1,m}(\xi)(1+\epsilon) \\
\theta_2(\epsilon,\xi)=\theta_{2,m}(\xi)(1+\epsilon)
\end{array}
\right. . \label{ThetaParam}
\end{equation}
Varying $\xi$ in the range $[-1,1]$ and $\epsilon$ in the range
$[-1,\infty]$, we can obviously cover the whole upper half of the
observer sky, since $\xi$ establishes the anomaly of the light ray
w.r.t. a reference axis in the sky (through Eqs.
(\ref{Th1m})-(\ref{Th2m})) and $\epsilon$ fixes the angular
distance from the center of the black hole.

In this paper, we are interested into light rays experiencing very
large deflections by a Kerr black hole. These light rays reach the
observer from directions $(\theta_1,\theta_2)$ very close to the
shadow. In the parametrization (\ref{ThetaParam}), they are thus
described by light rays with very small positive $\epsilon$, while
keeping $\xi$ in the whole range $[-1,1]$. The SDL amounts to
performing the integrals in the geodesics equations
(\ref{Geod1})-(\ref{Geod2}), to the lowest orders in $\epsilon$.

Now, we can easily derive the values of $J$ and $Q$ for these
strongly deflected photons using Eqs. (\ref{JTh1})-(\ref{QTh2}):
\begin{eqnarray}
&J(\xi,\epsilon)=& J_m(\xi) (1+\epsilon) \label{JSFL}\\
&Q(\xi,\epsilon)=& Q_m(\xi) (1+2\epsilon). \label{QSFL}
\end{eqnarray}

Substituting these expressions in Eq. (\ref{R}) and solving the
equation $R=0$ for $x_0$, we get the closest approach distance as
\begin{eqnarray}
&& x_0(\xi,\delta)= x_m(\xi)(1+\delta) \label{x0SFL} \\ && \delta
= \sqrt{\frac{2\epsilon}{3}}\left[ 1 -\frac{2}{3\sqrt{3}}a \,
\xi-\frac{8}{27}a^2(4\xi^2-3) \right] \label{deltatoeps}
\end{eqnarray}

In general, we see that the relation between $\delta$ and
$\epsilon$ depends on $\xi$, contrarily to what happens in the
Schwarzschild case, which, by the way, is correctly recovered when
$a$ is set to zero (compare with Ref. \cite{Boz1}). In the
resolution of the geodesics equation we will mostly use $\delta$
rather than $\epsilon$. However, they can be immediately
interchanged by Eq. (\ref{deltatoeps}) and its inverse.

\subsection{Resolution strategy}

Let us introduce our gravitational lensing configuration. As said
before, we restrict to observers on the equatorial plane of the
black hole at radial coordinate $D_{OL}$. We choose the zero of
the azimuthal Boyer-Lindquist coordinate $\phi$ on the observer
position. The source is assumed to be static at Boyer-Lindquist
coordinates $(D_{LS},\vartheta_s,\phi_s)$. To make contact with
previous works, we call $\gamma=\phi_s-\pi$.

Our lens equations are provided by Eqs.
(\ref{Geod1})-(\ref{Geod2}), where we identify $\phi_f=0$,
$\phi_0=\phi_s$. In these equations there are four different
integrals to solve:
\begin{eqnarray}
&& I_1= 2\int\limits_{x_0}^{\infty} \frac{dx}{\sqrt{R}} \label{I1}\\
&& I_2=2\int\limits_{x_0}^{\infty} \frac{x^{2}+a^{2}-a J}{\Delta
\sqrt{R}} dx \label{I2}\\
&& J_1=\pm \int \frac{1}{\sqrt{\Theta}} d \vartheta \label{J1} \\
&& J_2=\pm \int \frac{csc^2\vartheta}{\sqrt{\Theta}} d \vartheta .
\label{J2}
\end{eqnarray}
In the radial integrals $I_1$ and $I_2$ we have taken the higher
extrema to be infinite, because we assume $D_{OL},D_{LS} \gg 1$.
As the two integrands go to zero as $x^{-2}$ for $x\rightarrow
\infty$, the relative errors committed in this approximation are
of order $D_{OL}^{-1}$ and $D_{LS}^{-1}$ respectively. Moreover,
since the only inversion point in the radial motion is $x_0$, the
infalling pieces and the outgoing pieces of the integral are equal
and we can solve the sign ambiguity considering only the outgoing
pieces and putting a factor 2 in front of the integrals. The
radial integrals $I_1$ and $I_2$ can then be solved using the SDL
technique explained in Ref. \cite{Boz1}. In practice, considering
photons with minimum distance very close to some $x_m$, and
introducing the parametrizations (\ref{JSFL}), (\ref{QSFL}),
(\ref{x0SFL}), we can expand the integrals in terms of the
parameter $\delta$, introduced before. The leading terms diverge
logarithmically as $\delta$ goes to zero, while the
next-to-leading order terms are constants in $\delta$. The details
of this procedure are reported in Appendix A. Here we just rewrite
the final results
\begin{eqnarray}
& I_1=&- a_1 \log \delta+ b_1 \\
& I_2=&- a_2 \log \delta+ b_2,
\end{eqnarray}
where the coefficients $a_i$ and $b_i$ are functions of $a$ and
$\xi$. Their expansions to second order in $a$ are reported in
Appendix A.

As regards the angular integrals (\ref{J1})-(\ref{J2}), it is
convenient to introduce the new variable $\mu=\cos \vartheta$. The
final results, expanded to second order in $a$ are functions of
$\xi$ and the source position $\mu_s=\cos \vartheta_s$. They are
reported with a full derivation in Appendix B. We will just recall
them in the following sections when we need them.

Once all integrals are calculated, we have to solve Eqs.
(\ref{Geod1})-(\ref{Geod2}) in terms of the source coordinates
$(\gamma,\mu_s)$, so that they are in the lens map form
\begin{equation}
\left\{
\begin{array}{l}
\mu_s=\mu_s(\delta,\xi)  \\
\gamma=\gamma(\delta,\xi)
\end{array} \right. . \label{LensApp}
\end{equation}

Since all transformations from $(\gamma,\vartheta_s)$ to
$(\gamma,\mu_s)$ and from  $(\theta_1,\theta_2)$ to $(\delta,\xi)$
are non-singular (except for the points $\xi=\pm 1$), the
singularities of the Jacobian of the map (\ref{LensApp}) represent
gravitational lensing critical points.

In the following sections, we will calculate the critical curves
and the caustics of the Kerr gravitational lens order by order.
Then we will describe the lens mapping in a neighborhood of a
generic caustic, deriving the images position and magnification.

\section{Derivation of the relativistic caustics}

\subsection{Zero order caustics}

Sending $a$ to zero, we must recover the Schwarzschild results,
i.e. that critical curves are concentric rings corresponding to
point-like caustics aligned on the optical axis, alternatively
located behind and in front of the black hole. Of course, as
$a\rightarrow 0$, all expressions are considerably simplified, and
it is possible to follow calculations without too much effort.

Reading all the zero-order results for the integrals in Appendices
A and B, Eq. (\ref{Geod1}) becomes

\begin{eqnarray}
& -2\log \delta& +2\log[12(2-\sqrt{3})] = \nonumber \\ && =\pm
\arcsin \frac{\mu_s}{\sqrt{1-\xi^2}}+m\pi. \label{PreEqmu0}
\end{eqnarray}

Defining the new variable
\begin{equation}
\psi=-2\log \delta +2\log[12(2-\sqrt{3})],\label{psidef}
\end{equation}
Eq. (\ref{PreEqmu0}) can be easily solved as
\begin{equation}
\mu_s=\pm \sqrt{1-\xi^2}\sin \psi. \label{Eqmu0}
\end{equation}

The second lens equation (\ref{Geod2}) now reads
\begin{equation}
\gamma =-(m-1)\pi \mp\arctan \frac{\mu_s
\xi}{\sqrt{1-\mu_s^2-\xi^2}}.
\end{equation}

Using Eq. (\ref{Eqmu0}) to eliminate $\mu_s$, we find
\begin{equation}
\gamma =-(m-1)\pi -\arctan \left[\xi \tan \psi \right].
\label{Eqgamma0}
\end{equation}

The number $m$ appearing in this equation is the integer part of
$(\psi/\pi+1/2)$ and must be interpreted as the number of
inversions in the polar motion of the photon.

Eqs. (\ref{Eqmu0}) and (\ref{Eqgamma0}) represent the lens
equations for a Schwarzschild black hole without the classical
identification of the equatorial plane with the
source-lens-observer plane. We can recover the results of Ref.
\cite{Boz1} imposing that the motion takes place on the equatorial
plane, i.e. setting $\xi=1$. Then we have $\mu_s=0$ (the source
must coherently lie on the equatorial plane) and
$\gamma=-\psi+\pi$. The quantity $\psi-\pi$ represents the
deflection angle of a photon approaching the Schwarzschild black
hole at a distance $x_0=x_m(1+\delta)$. Eqs. (\ref{Eqmu0}) and
(\ref{Eqgamma0}) can also be obtained from the traditional planar
treatment by a trivial rotation by an angle $\arccos \xi$ of the
reference plane around the optical axis using some spherical
trigonometry.

Now we can easily calculate the Jacobian of our lens map. We find
\begin{eqnarray}
&& \frac{\partial \mu_s}{\partial \xi}= \mp \frac{\xi}{\sqrt{1-\xi^2}}\sin \psi\\
&& \frac{\partial \mu_s}{\partial \psi}= \pm \sqrt{1-\xi^2}\cos \psi\\
&& \frac{\partial \gamma}{\partial \xi}= -\frac{ \tan
\psi}{1+\xi^2 \tan^2 \psi} \\ && \frac{\partial \gamma}{\partial
\psi}= -\frac{\xi \sec^2 \psi}{1+\xi^2 \tan^2 \psi}
\end{eqnarray}
and
\begin{equation}
D= \frac{\partial \mu_s}{\partial \xi} \frac{\partial
\gamma}{\partial \psi}-\frac{\partial \mu_s}{\partial
\psi}\frac{\partial \gamma}{\partial \xi}= \pm \frac{\sin
\psi}{\sqrt{1-\xi^2}}.
\end{equation}

The critical curves are the solutions of the equation $D=0$,
which, in our case, simply gives
\begin{equation}
\psi_{cr}= k \pi. \label{CC0}
\end{equation}

The critical $\psi$ does not depend on $\xi$. Recalling that
$\psi$ is just a function of $\delta$ expressed by Eq.
(\ref{psidef}), we have
\begin{equation}
\delta_{cr}=12(2-\sqrt{3})e^{-k\pi/2}. \label{deltacrit}
\end{equation}
Switching to $\epsilon$ by Eq. (\ref{deltatoeps}), we have
\begin{equation}
\epsilon_{cr}=216(2-\sqrt{3})^2e^{-k\pi}. \label{epscrit0}
\end{equation}
Then, recalling the meaning of $\epsilon$ by Eq.
(\ref{ThetaParam}) and taking $\theta_{1,m}$ and $\theta_{2,m}$
from Eqs. (\ref{Th1m})-(\ref{Th2m}), we finally find
\begin{equation}
\begin{array}{l}
D_{OL} \theta_{1,cr}(\xi)= -\frac{3 \sqrt{3} \xi}{2} \left[1+ \epsilon_{cr}(k) \right]\\
D_{OL} \theta_{2,cr}(\xi) = \pm  \frac{3 \sqrt{3}}{2}
\sqrt{1-\xi^2}\left[1+ \epsilon_{cr}(k) \right]
\end{array}, \label{Crit0}
\end{equation}
that is a series of rings parameterized by $\xi$, slightly larger
than the shadow of the black hole, in full agreement with Refs.
\cite{Dar,Nem,BCIS,Boz1}. The critical curves are labelled by the
number $k$. We shall refer to $k$ as the critical curve order or
the caustic order, when we consider the corresponding caustic.

Coming to the caustics, inserting Eq. (\ref{CC0}) in the lens
equations Eqs. (\ref{Eqmu0})-(\ref{Eqgamma0}) and noting that the
number of inversions in polar motion is $m=k$, we find
\begin{equation}
\mu_s=0, ~~ \gamma=-(k-1)\pi.
\end{equation}

The caustics are points aligned on the optical axis. For odd $k$
they are behind the black hole, while for even $k$ they are in
front of the black hole (retro-lensing caustics). In the
Schwarzschild limit, the number of loops performed by photons
forming critical images of order $k$ is just $(k-1)/2$. However,
this is not exactly true for spinning black holes, as we shall see
in the next section. So, it is better to think of the order of the
critical curve $k$ as the number of inversions in the polar motion
done by photons associated to it.

Of course, the first caustic for $k=1$ is the classical weak field
limit one, which is outside the range of the SDL approximation, so
we cannot expect to recover the Einstein ring radius putting $k=1$
into Eq. (\ref{Crit0}) (the first caustic is no longer described
even by the weak field approximation if the source is close to the
black hole). However, as shown in Refs. \cite{Boz1,S1-S14}, the
SDL approximation works better and better for higher $k$, starting
from the first retro-lensing caustic in $k=2$. It is these
caustics that we are going to study in the following sections. In
particular, we will find out how they are displaced and deformed
by the black hole spin, obtaining a full analytic description of
their shape.

\subsection{First order caustics}

Up to now we have just re-obtained all the Schwarzschild black
hole results in a more complicated form, starting from Kerr
geodesics equations and sending $a$ back to zero. Now, we shall
introduce first order corrections to our lens equations,
re-deriving the critical curves and the caustics. We anticipate
that the caustics get displaced from the optical axis in the
azimuthal direction, though remaining point-like.

Using the first order terms of the radial and angular integrals
from the Appendices A and B, we can add the terms proportional to
$a$ in the equations (\ref{Eqmu0}), (\ref{Eqgamma0}). The
inversion of the $\mu_s$ equation can be easily performed order by
order in $a$. Then, repeating the same steps of the previous
section, we find
\begin{equation}
\mu_s= \pm \sqrt{1-\xi^2}\sin \psi \pm \frac{4a \xi}{3\sqrt{3}}
\sqrt{1-\xi^2}\sin \psi\label{Eqmu1}
\end{equation}

\begin{eqnarray}
& \gamma =& -(m-1)\pi -\arctan \left[\xi \tan \psi \right]
\nonumber \\ && -\frac{4a}{3\sqrt{3}} \left[ \psi+\frac{\xi^2\tan
\psi}{1-(1-\xi^2)\sin^2\psi} \right. \nonumber \\ && \left. -\tan
\psi +3\sqrt{3} \log(2\sqrt{3}-3) \right].
 \label{Eqgamma1}
\end{eqnarray}

Note that for $\psi$ close to $m\pi+\pi/2$ and $\xi$ close to
zero, the first order correction to $\mu_s$ may bring it to
absolute values larger than 1. As $\mu_s$ is the cosine of the
polar angle, these values are unphysical. This inconsistency comes
out because, when we solve for $\mu_s$ order by order, we expand
the $\arcsin$ function in points very close to the extrema of its
definition range, where the $\arcsin$ is not analytic. Then, the
linear approximation obviously takes us out of the interval
$[-1,1]$. The dangerous values of $\psi$ and $\xi$ correspond to
nearly polar trajectories where the final direction is very close
to one of the two poles. However, as we shall see, the highest
magnification for the relativistic images is obtained when the
source is close to a caustic. Luckily, the caustics lie at $\mu
\simeq 0$ for an equatorial observer, so that we will always work
very far from these dangerous regions. This danger will become
effective for very high order caustics, which may become very
large, as we shall see in the next subsection.

Now we can calculate the derivatives of the lens equation as
before. The Jacobian reads
\begin{equation}
D= \pm \frac{\sin \psi}{\sqrt{1-\xi^2}}
\left(1+\frac{8a\xi}{3\sqrt{3}} \right),
\end{equation}
which tells us that the critical curves are still described by Eq.
(\ref{CC0}) even at first order. This means that there is no
correction to the critical $\delta$ (\ref{deltacrit}) of the
previous section. The fact that we do not get any corrections to
$\delta$ does not mean that the shape of the critical curves is
not altered by the black hole spin at first order. Indeed going
back from $\delta$ to $\epsilon$ we get a first order correction,
according to Eq. (\ref{deltatoeps}). Moreover, the shadow shape is
modified according to Eqs. (\ref{Th1m})-(\ref{Th2m}). Then, to
first order, the critical curves are
\begin{equation}
\begin{array}{ll}
D_{OL} \theta_{1,cr}(\xi)=& -\frac{3 \sqrt{3} \xi}{2}\left(1+\epsilon_{cr} \right) \\
& +2a\left[1+ \epsilon_{cr}(1-\xi^2) \right] \\
D_{OL} \theta_{2,cr}(\xi) =&  \pm  \frac{3 \sqrt{3}}{2}
\sqrt{1-\xi^2}\left[1+ \epsilon_{cr} \right] \\
& \pm 2a\epsilon_{cr}\xi \sqrt{1-\xi^2}
\end{array}, \label{Crit1}
\end{equation}
with $\epsilon_{cr}$ still given by Eq. (\ref{epscrit0}).

Also the caustics are modified. In fact, plugging $\psi_{cr}=k\pi$
into Eqs. (\ref{Eqmu1})-(\ref{Eqgamma1}), we find
\begin{eqnarray}
&& \mu_s=0 \\ && \gamma =-(k-1)\pi -4a \left[ \frac{k
\pi}{3\sqrt{3}}+ \log(2\sqrt{3}-3) \right] \label{Cau1}
\end{eqnarray}

The caustics are still fully confined to the equatorial plane,
they are still point-like, but they are displaced from the optical
axis. The displacement is negative, which means that the caustics
drift clockwise if we look at the Kerr black hole from the
northern pole. So, we can say that if a source lies on a caustic
point of order $k$, prograde light rays perform more than
$(k-1)/2$ loops while retrograde light rays perform less than
$(k-1)/2$ loops. The number of inversions in the polar motion is
still $k$. Higher order caustics are more displaced, because of
the $k$ dependence in Eq. (\ref{Cau1}). Of course, as said before,
this formula correctly describes all caustics except for the weak
field one, corresponding to $k=1$. Going to second order in $a$ we
will describe the full shape of the caustics.

\subsection{Second Order Caustics}

At first order in $a$ the caustics still remain point-like, while
it is known that they get a finite extension when $a$ is different
from zero \cite{RauBla,BozEq}. So, it is necessary to proceed to
second order. To the right hand side of Eqs.
(\ref{Eqmu1})-(\ref{Eqgamma1}) we have to add the following
quantities respectively:
\begin{eqnarray}
&\delta\mu_s^{(2)}&=\pm \frac{\sqrt{1-\xi^2}}{108\sqrt{3}}a^2
\times \nonumber \\ && \left[ 12\sqrt{3}(1-\xi^2)\psi_n\cos
\psi-\sqrt{3}(1+31\xi^2)\sin \psi \right. \nonumber \\ && \left.
-\sqrt{3}(1-\xi^2)\sin 3\psi \right] \label{dmu2}
\end{eqnarray}
\begin{eqnarray}
&\delta\gamma^{(2)}&=\frac{2a^2}{27\xi^2}(1-\xi^2)(23\xi^2-4)\arctan(\xi
t) \nonumber \\ &&
+\frac{2}{9}a^2\xi(16-\psi_n)-\frac{1-\xi^2}{9(1+\xi^2t^2)}a^2\xi
 \left[ (1+t^2) \psi_n \right. \nonumber \\ && \left.
+\frac{t}{3}\left(23-\frac{8}{\xi^2}-16 \frac{1+t^2}{1+\xi^2 t^2}
\right) \right] \label{dgamma2}
\end{eqnarray}
where $t=\tan \psi$ and $\psi_n=5\psi+8\sqrt{3}-20$.

Note that a term directly proportional to $\psi$ appears in Eq.
(\ref{dmu2}) through $\psi_n$. This is another danger for the
approximation, since for very large $\psi$, i.e. photons
performing several loops around the black hole, $\delta
\mu_s^{(2)}$ may become even larger than 1. This break-down sets
the true limit to the applicability range of the perturbative
expansion in $a$, which gets smaller and smaller for photons
making many loops. However, the brightest images are formed by
photons associated to critical images of low order. For these
images, as we shall see, the range of applicability of the
perturbative expansion is considerably large.

Now let us find out the corrections to the critical curves. Once
we have calculated all derivatives of the lens equations and
written the Jacobian to the second order in $a$, we set
\begin{equation}
\psi=k\pi+a^2 \delta\psi, \label{psipert}
\end{equation}
as we already know the zero order critical curve and we know that
there is no correction at the first order. Then we easily get rid
of all the trigonometric functions and the final Jacobian reads
\begin{equation}
D=\pm \frac{(-1)^k a^2}{9\sqrt{1-\xi^2}}\left[9\, \delta\psi-
(92-24\sqrt{3}-15k\pi)(1-\xi^2) \right]. \label{Jac2}
\end{equation}

The $(-1)^k$ is a consequence of the expansions of the
trigonometric functions, while the double sign inherited by the
Jacobian at all orders depends on the fact that the
$(\epsilon,\xi)$ parametrization only covers half of the observer
sky and we are forced to introduce a double sign in the equation
for $\mu_s$.

The equation $D=0$ gives the second order correction to the
critical $\psi$ in a very simple form
\begin{equation}
\delta\psi=\frac{92-24\sqrt{3}-15k\pi}{9}(1-\xi^2). \label{dpsi}
\end{equation}

Now we can remount the complete second order expansion of the
critical curves, which reads
\begin{equation}
\begin{array}{ll}
D_{OL} \theta_{1,cr}(\xi)=& -\frac{3 \sqrt{3} \xi}{2}\left(1+\epsilon_{cr} \right) \\
& +2a\left[1+ \epsilon_{cr}(1-\xi^2) \right] \\
& -\frac{\xi a^2}{\sqrt{3}} \left\{ \xi^2-2 +\left[ \frac{15}{2}k\pi(1-\xi^2) \right. \right. \\
& \left. \left. +\frac{176}{3}-12\sqrt{3}+ \left( \frac{179}{3}-12\sqrt{3} \right) \xi^2 \right] \epsilon_{cr} \right\}\\
D_{OL} \theta_{2,cr}(\xi) =&  \pm  \frac{3 \sqrt{3}}{2}
\sqrt{1-\xi^2}\left[1+ \epsilon_{cr} \right] \\
& \pm 2a\epsilon_{cr}\xi \sqrt{1-\xi^2} \\
& \pm \frac{\sqrt{1-\xi^2}}{\sqrt{3}} a^2\left\{ \xi^2+ \left[
\frac{15}{2}k\pi(1-\xi^2)\right.\right. \\
& \left. \left. + 12\sqrt{3}-54 + \left(\frac{179}{3}-12\sqrt{3}
\right) \xi^2 \right] \epsilon_{cr} \right\}
\end{array}. \label{Crit2}
\end{equation}

Here, again, we have $k\pi$ terms which become large for higher
order critical curves.

Finally, let us calculate the caustics at the second order in $a$.
Plugging Eq. (\ref{psipert}) with (\ref{dpsi}) into the lens
equations, we find
\begin{eqnarray}
& \mu_s=& \pm (-1)^k r_c (1-\xi^2)^{3/2} \label{Caumu}\\ & \gamma
=&-(k-1)\pi -4a \left[ \frac{k \pi}{3\sqrt{3}}+ \log(2\sqrt{3}-3)
\right] \nonumber \\ && + r_c \xi^3,\label{Caugamma}
\end{eqnarray}
where we define
\begin{equation}
r_c=\frac{2}{9}a^2(5k\pi+8\sqrt{3}-36). \label{rc}
\end{equation}

\begin{figure}
\resizebox{\hsize}{!}{\includegraphics{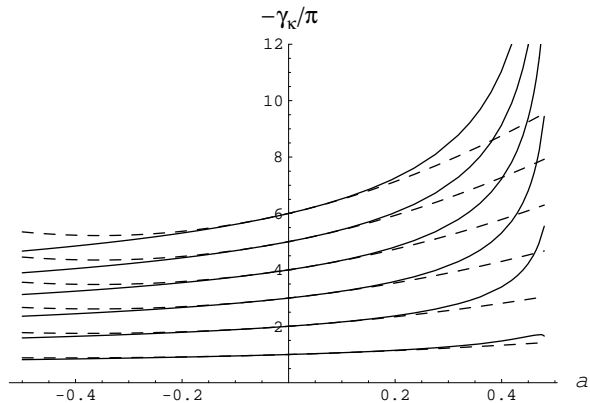}} \caption{
Comparison between the intersections of the caustics with the
equatorial plane as calculated in Ref. \cite{BozEq} without the
perturbative approximation for the black hole spin (solid lines)
and the ones calculated in the present paper (dashed lines). The
plot refers to the caustics of order $2\leq k\leq 7$.
 }
 \label{Fig gammak}
\end{figure}

The analytical expressions of the Kerr black hole caustics, given
by Eqs. (\ref{Caumu})-(\ref{Caugamma}) to the second order in the
black hole spin $a$, represent a major achievement of this paper.
Before discussing their shape and all the physical implications,
it is a good idea to test our formulae by comparing them to the
results obtained by alternative methods. In Ref. \cite{BozEq}, the
intersections of the caustics with the equatorial plane were found
using the SDL approximation only, without any limitation for the
black hole spin. The first test is to analytically expand the
formulae of Ref. \cite{BozEq} to the second order in $a$, without
using any numerical integrations. Indeed, we get the same
expressions as in Eq. (\ref{Caugamma}), evaluated for $\xi=\pm 1$.
Furthermore, we can draw in the same plot the intersections of the
caustics with the equatorial plane as calculated in this paper
along with those calculated in Ref. \cite{BozEq}. Rather than
making two separate plots for prograde and retrograde photons, we
can make a unique plot, letting $a$ vary in the range $[-0.5,0.5]$
and keeping the values of Eq. (\ref{Caugamma}) for $\xi=1$. So,
the left side of the plot ($a<0$) represents the intersections for
retrograde photons and the right side $(a>0)$ represents the
intersections of prograde photons. We see in Fig. \ref{Fig gammak}
that the second order approximations (dashed lines) follow the
exact expressions of Ref. \cite{BozEq} very accurately. We can
estimate that for lower order caustics the perturbative
approximation works up to $a\simeq 0.3$, while for the last
caustic in the plot ($k=7$) we have to stop at $a\simeq 0.1$. In
any case, the validity range is impressively large, reaching
values of the black hole spin comparable to the extremal case.
This encourages us to make extensive and confident use of the
second order approximation for a full description of Kerr lensing
phenomenology.

\begin{figure}
\resizebox{\hsize}{!}{\includegraphics{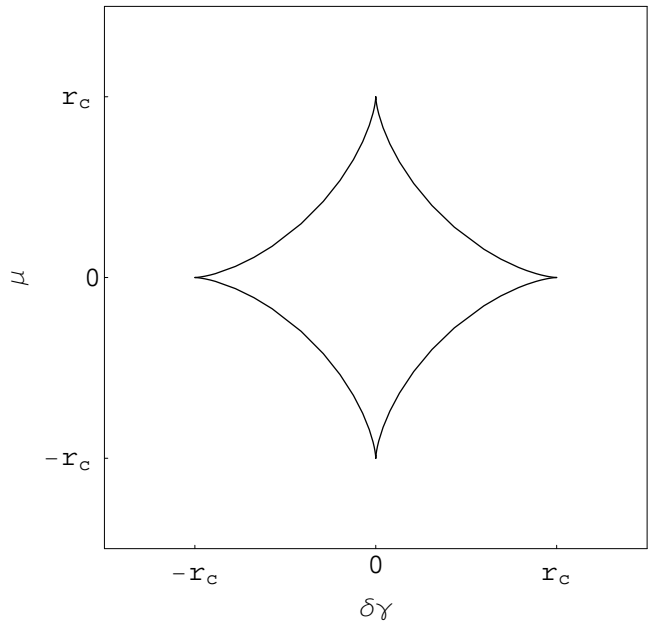}} \caption{ The
typical caustic in Kerr gravitational lensing has the astroid
shape and the same angular extension $r_c$ (given by Eq.
(\ref{rc})) along the azimuthal and the polar direction.}
 \label{Fig Caus}
\end{figure}

Now, let us discuss the shape and the extension of Kerr lensing
caustics. Looking at Eq. (\ref{Caumu}), it is interesting to note
that the upper half of the critical curve is mapped in the lower
half of the caustic for odd $k$, while it is mapped in the upper
half if $k$ is even. As already found numerically in Ref.
\cite{RauBla}, the caustics have the characteristic astroid shape
shown in Fig. \ref{Fig Caus}, common to all tangential caustics
after the breaking of the axial symmetry. The four cusps are in
$\xi=\pm 1$ and $\xi=0$ choosing different signs for $\mu_s$.

The caustics have the same extension in $\gamma$ and $\mu$. We
recall that $\gamma$ is just the azimuthal angle of the
Boyer-Lindquist coordinates taken from the reference axis starting
from the black hole and going in the direction opposite to the
observer, while $\mu$ is the cosine of the polar angle
$\vartheta$. As said before, we can trust our results as long as
the perturbative terms remain small. In this regime, $\mu \simeq
\frac{\pi}{2}-\vartheta$. Then, we deduce that the caustics have
the same extension in the azimuthal and in the polar direction.

The extension is parameterized by the quantity $r_c$, which is the
semi-axis of the caustic. We see that it grows with the black hole
spin $a$ and with the caustic order $k$. Thus, for higher caustic
orders, the perturbative approximation fails for smaller and
smaller values of $a$, while it remains good for lower order
caustics. This was already noted while commenting Fig. \ref{Fig
gammak}.

\begin{table}
\begin{tabular}{c|cccc}
  \hline
  & a=0.01 & a=0.05 & a=0.1 & a=0.2 \\
  \hline
  $\Delta \gamma_2$ & 0.018 & 0.088 & 0.18 & 0.35 \\
  $\Delta \gamma_3$  & 0.042 & 0.21 & 0.42 & 0.84 \\
  $\Delta \gamma_4$ & 0.066 & 0.33 & 0.66 & 1.3 \\
  $\Delta \gamma_5$  & 0.09 & 0.45 & 0.9 & 1.8 \\
  $\Delta \gamma_6$ & 0.11 & 0.57 & 1.14 & 2.3 \\
  $\Delta \gamma_7$  & 0.14 & 0.69 & 1.39 & 2.8 \\
  \hline
\end{tabular}
\caption{Drift (in radians) of the caustics of order $k$ with
$2\leq k\leq 7$, for different values of $a$.}
\end{table}

The drift from the optical axis of the caustic is given by the
first order term in Eq. (\ref{Caugamma}). Indicating it by $\Delta
\gamma$, in Tab. 1 we report the obtained values for the first 6
relativistic caustics, starting from $k=2$, for values of the
black hole spin going up to $a=0.2$. We see that the drift may
become very large, amounting to several radians already for the
5th order caustic, while still in a regime where the perturbative
solution is valid. For higher order caustics, the number of loops
may be significantly different from the planar orbit result
$(k-1)/2$. Another important consideration is that we do not need
perfect alignment between source, lens and observer to have
gravitational lensing. To enhance the images associated with a
critical curve of order $k$, the source must align with the
corresponding order $k$ caustic, which may be well off the optical
axis. Moreover, the relativistic images are not enhanced all at at
the same time, since caustics of different order move far away
each other.

\begin{table}
\begin{tabular}{c|cccc}
  \hline
  & a=0.01 & a=0.05 & a=0.1 & a=0.2 \\
  \hline
  $r_{c,2}$ & 0.00021 & 0.0052 & 0.021 & 0.082 \\
  $r_{c,3}$ & 0.00056 & 0.014 & 0.056 & 0.22 \\
  $r_{c,4}$ & 0.0009 & 0.023 & 0.09 & 0.36 \\
  $r_{c,5}$ & 0.0013 & 0.031 & 0.13 & 0.5 \\
  $r_{c,6}$ & 0.0016 & 0.04 & 0.16 & 0.64 \\
  $r_{c,7}$ & 0.002 & 0.05 & 0.2 & 0.78 \\
  \hline
\end{tabular}
\caption{Radius (in radians) of the caustics of order $k$ with
$2\leq k\leq 7$, for different values of $a$.}
\end{table}

In Table 2 we report the radii of the first 6 relativistic
caustics for different values of $a$. The extension of the
caustics is of second order in $a$ and thus remains much smaller
than the drift, reaching some tenth of radians in the perturbative
regime. Outside of this regime, it is difficult to know what would
happen to higher order caustics. Would their vertical extension
saturate before reaching the poles or would they wrap around the
pole? Would they meet each other and make transitions to more
complicated structures? The answers to these questions need
further research, both analytically and numerically. We just want
to remark that the finite extension of relativistic caustics is of
striking importance for phenomenology, as will be clear in the
next sections.

\section{Gravitational lensing near caustics}

The description of the caustics is the fundamental step for a full
description of gravitational lensing. In this section we will give
a complete analytic resolution of the Kerr lens equation for
sources close to relativistic caustics.

The starting point is the second order lens equations, built
adding (\ref{dmu2}) to (\ref{Eqmu1}) and (\ref{dgamma2}) to
(\ref{Eqgamma1}). Let us consider a source whose distance from the
$k$-th order caustic is of order $a^2$ (thus being comparable to
the caustic size). Then its position can be expressed in the
following way
\begin{eqnarray}
&& \mu_s=a^2 \delta\mu_s \label{mua2}\\ &&
\gamma=\gamma_{cau}^{(1)}+a^2 \delta\gamma, \label{gammaa2}
\end{eqnarray}
where $\gamma_{cau}^{(1)}$ is the caustic position at the first
order in $a$, expressed by Eq. (\ref{Cau1}). Correspondingly, the
images associated to the critical curve of order $k$ will be
enhanced. They will be described by
\begin{eqnarray}
\psi=k\pi + a^2 \delta \psi. \label{psia2}
\end{eqnarray}

Substituting Eqs. (\ref{mua2})-(\ref{psia2}) in the lens
equations, the zero and first order terms cancel out and we are
only left with the second order terms
\begin{eqnarray}
&\delta\mu_s=& S \frac{(-1)^k}{9} \sqrt{1-\xi^2} \left[ 9\, \delta
\psi \right. \nonumber \\ && \left. +(1-\xi^2)(5k\pi+
8\sqrt{3}-20) \right] \label{Lensmu} \\ &\delta \gamma=&
-\frac{\xi}{9} \left[ 9\, \delta \psi-32 \right. \nonumber \\ &&
\left. +(3-\xi^2)(5k\pi+ 8\sqrt{3}-20) \right].\label{Lensgamma}
\end{eqnarray}

This is the Kerr lens equation close to the caustic of order $k$.
$S$ is just a sign which takes into account the fact that the
$(\psi,\xi)$ parametrization only covers half of the observer sky.
So, $S=+1$ for the upper half of the observer sky and $S=-1$ for
the lower half. We can easily check that the Jacobian (\ref{Jac2})
is just $a^2$ times the Jacobian of this lens equation
\begin{equation}
\frac{D}{a^2}= \frac{\partial (\delta \mu_s)}{\partial \xi}
\frac{\partial (\delta\gamma)}{\partial (\delta
\psi)}-\frac{\partial (\delta \mu_s)}{\partial
(\delta\psi)}\frac{\partial (\delta\gamma)}{\partial \xi}.
\end{equation}

The surprisingly simple form of the lens equation encourages its
analytical resolution. The $\delta \gamma$ equation can be easily
solved for $\delta \psi$:
\begin{equation}
\delta \psi=- \frac{\delta \gamma}{\xi} +\frac{1}{9}\left[ 32
-(3-\xi^2)(5k\pi+ 8\sqrt{3}-20) \right]. \label{Eqpsiimages}
\end{equation}
Plugging this expression into the $\delta \mu_s$ equation, we can
write it in the form
\begin{equation}
\delta\mu_s \xi= -S (-1)^k\sqrt{1-\xi^2}(\delta\gamma+ x_c \xi),
\label{Eqimages}
\end{equation}
where $x_c=r_c/a^2$ and $r_c$ is the semi-axis of the caustic as
defined by Eq. (\ref{rc}). Squaring both sides we get a fourth
order equation for $\xi$
\begin{equation}
x_c^2\xi^4+2x_c \delta\gamma \xi^3+(\delta \gamma^2 +\delta
\mu_s^2-x_c^2)\xi^2-2x_c \delta\gamma \xi -\delta \gamma^2=0.
\label{Eqxi4}
\end{equation}

The real solutions of this equation are images for a source in
$(\gamma_{cau}^{(1)}+a^2 \delta\gamma,a^2\delta \mu_s)$. It is
easy to check that we have two images if the source is outside the
caustic and four images if it is inside.

Once we have found the solutions of the squared equation, we have
to go back to the original equation (\ref{Eqimages}). Each root of
Eq. (\ref{Eqxi4}) satisfies Eq. (\ref{Eqimages}) only with one
choice of $S$. This determines the half-sky where the image
appears. It is the upper half if $S=1$ and the lower half if
$S=-1$. Then, we can easily calculate the value of $\delta \psi$
for each image through (\ref{Eqpsiimages}) and then go back to
$\epsilon$ by Eqs. (\ref{psidef}) and (\ref{deltatoeps}).  Finally
we can write the images as
\begin{eqnarray}
&D_{OL}\theta_1&= -\frac{3\sqrt{3}}{2}\xi(1+\epsilon_{cr})+
2a[1+\epsilon_{cr}(1-\xi^2)] \nonumber \\ && \!\!\!\!\!\!\!
+\frac{a^2\xi}{6\sqrt{3}}
[12-6\xi^2+\epsilon_{cr}(76-82\xi^2+27\delta \psi)]
\label{ImagePos1} \\ &D_{OL} \theta_2&= S \frac{3\sqrt{3}}{2}
\sqrt{1-\xi^2} \left[ (1+\epsilon_{cr})+\frac{4}{3\sqrt{3}} a\xi
\epsilon_{cr} \right. \nonumber
\\
&& \!\!\!\!\!\!\! \left.
+\frac{a^2}{27}(6\xi^2-\epsilon_{cr}(48-82\xi^2+27 \delta \psi))
\right],  \label{ImagePos2}
\end{eqnarray}
with $\xi$ and $\delta \psi$ solving Eqs.
(\ref{Lensmu})-(\ref{Lensgamma}).

In the particular case $\delta \mu_s =0$ (source on the equatorial
plane), the solutions are $\xi=\pm 1$ and $\xi=\delta \gamma/x_c$
(double root). The first two solutions are two images staying on
the equatorial plane, which are physical for any value of $\delta
\gamma$. The other two are acceptable only if $|\delta \gamma|
\leq x_c$ because $\xi$ is defined in the range $[-1,1]$. This is
in agreement with the fact that $x_c$ represents the caustic
semi-axis. These two images form symmetrically w.r.t. the
equatorial plane, grazing the critical curve from the outside.
When these additional images are present, the former two are
inside the critical curve, while when they are absent the
remaining images are one inside and the other outside of the
critical curve so that global parity is conserved in caustic
crossing.

On the other hand, if $\delta \gamma=0$, we have the solutions
$\xi=0$ (double root) and $\xi=\pm \sqrt{1-(\delta \mu_s/x_c)^2}$.
The first two form very close to the polar direction on opposite
sides of the critical curve, while the last two are real only for
$|\delta \mu_s|<x_c$. They form symmetrically w.r.t. the polar
direction and graze the critical curve from the inside. As before,
things work in such a way that global parity is conserved.

\begin{figure}
\resizebox{\hsize}{!}{\includegraphics{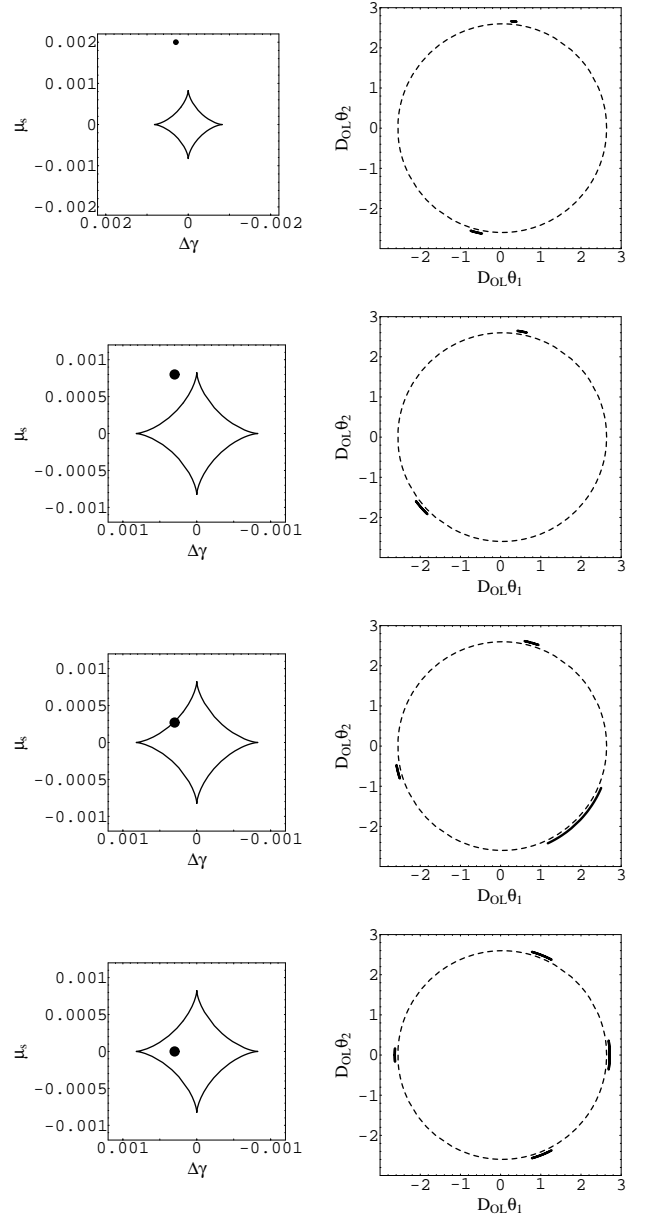}} \caption{
Formation of the images for a source approaching the first
restrolensing caustic ($k=2$). On the left we show several
positions for a source and on the right we have the corresponding
images around the shadow (in dashed style). The thickness of the
images has been exaggerated to make them more evident.}
 \label{Fig k=2}
\end{figure}

Now let us make an example with a physical source and a physical
black hole. Sgr A* has a mass of $3.61\times 10^5$ solar masses.
Let us suppose that its spin is $a=0.02$ (Liu \& Melia estimate
$a=0.044$ \cite{LiuMel}, but different methods point to different
values). Then we are able to calculate the caustic positions and
shapes. As a source, consider a star with a radius $R_S=3R_\odot$
at $200AU$ from Sgr A*. This is the order of magnitude of the
periapse distance of the observed stars orbiting Sgr A*, like S2
or S14 \cite{Eis}. In Fig. \ref{Fig k=2} we show what we would see
if this star approaches the first retro-lensing caustic. The
position of this caustic is in $\gamma=177.98^\circ$, so slightly
displaced from the optical axis. This means that the source should
be almost in front of the observer, very close to the optical
axis. On the left panels of Fig. \ref{Fig k=2} we show several
positions of the source relative to the caustic, as seen by the
black hole. Notice that with these values of $a$, source radius
and distance the size of the source is comparable to that of the
caustic. On the right panels, we show the corresponding images and
the shadow in dashed style. We see that when the source is far
from the caustic (top panels) there are two small images. The
bigger one is below the black hole if the source is above the
equatorial plane (we recall that this is normal in a retro-lensing
situation). When the source approaches the caustic (second row
panels), the two images do not lie on opposite sides w.r.t the
black hole. This distortion is a consequence of the axial symmetry
breaking. When the source enters the caustic, two more images form
(third and fourth row panels). If the source orbits with a
velocity of the order of the circular orbit speed $\sqrt{G
M/D_{LS}}$, the whole caustic crossing takes 3.4 hours, much
longer than the typical times of the primary caustic crossing,
which takes just few seconds \cite{RauBla}. Furthermore, since the
higher order caustics are much more extended, the probability of
caustic crossing is much higher.

\begin{figure}
\resizebox{\hsize}{!}{\includegraphics{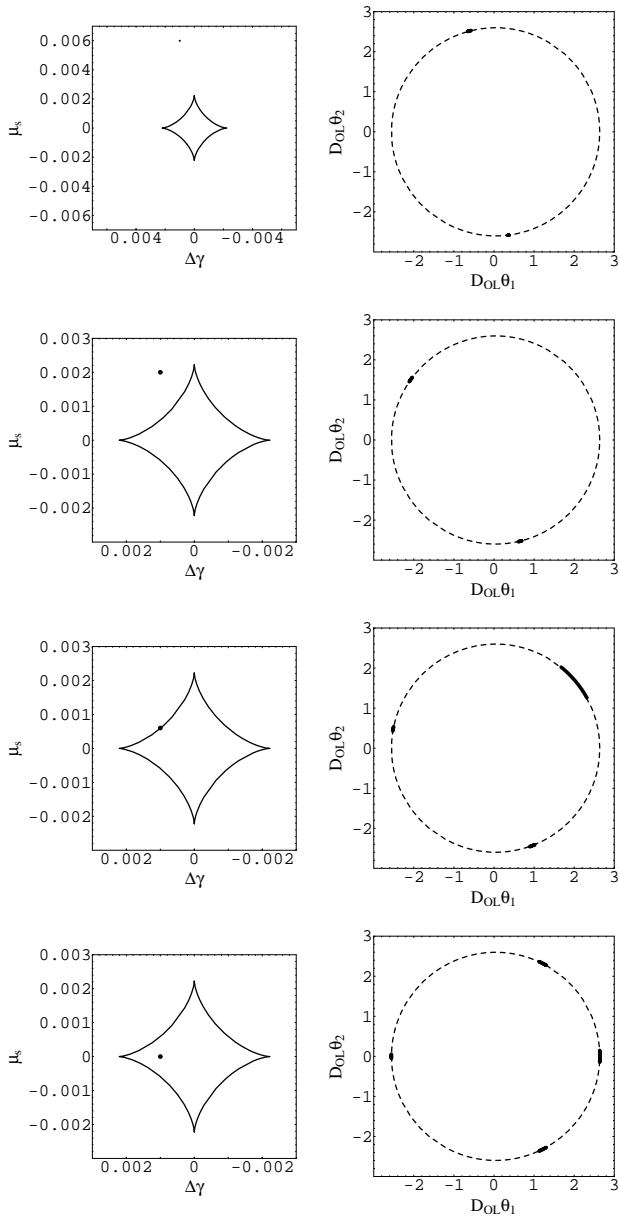}} \caption{
Formation of the images for a source approaching the first
relativistic standard lensing caustic ($k=3$). On the left we show
several positions for a source and on the right we have the
corresponding images around the shadow (in dashed style). The
thickness of the images has been exaggerated to make them more
evident.}
 \label{Fig k=3}
\end{figure}

In Fig. \ref{Fig k=3} we have shown the case where the same source
approaches the first relativistic standard lensing caustic
($k=3$), which now is displaced to $\gamma=4.8^\circ$ on the right
of the black hole. As this caustic is larger, the source now looks
smaller compared to the caustic, as we see in left panels. When
the source is far from the caustic (top panels), there are two
images, the bigger one being on the same side of the source
(standard lensing situation). As the sources approaches, the
images and the black hole are no longer on the same line (second
row panels), then formation of two new images takes place (third
and fourth row panels). In this case, the caustic crossing takes
9.2 hours, for a source velocity equal to the circular orbit
velocity.

\begin{figure}
\resizebox{\hsize}{!}{\includegraphics{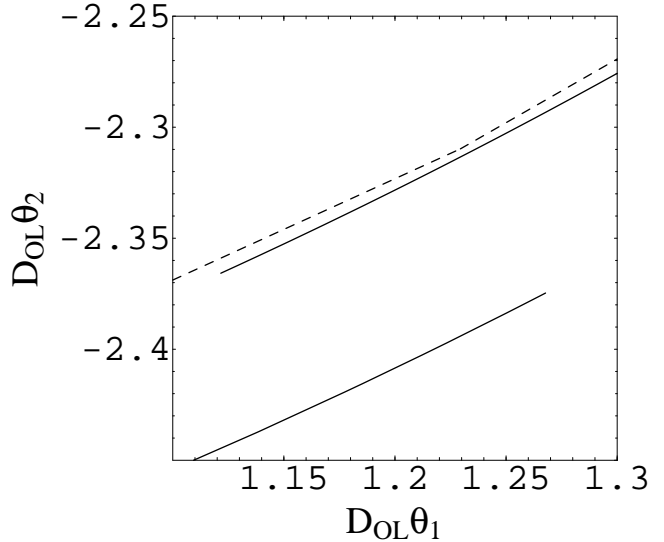}}
\caption{A zoom very close to the shadow border (in dashed style),
showing at the same time the images of two sources, one being in
the $k=2$ caustic (outer tangential arc) and the other being in
the $k=3$ caustic (inner tangential arc).}
 \label{Fig Compk23}
\end{figure}

Fig. \ref{Fig Compk23} zooms on two images generated by different
sources in the caustics $k=2$ and $k=3$. The first source
generates the outer tangential arc while the second source
generates the inner tangential arc. This is because the more loops
the photons perform, the closer they get to the black hole. Then
higher order images appear closer and closer to the shadow. What
astonishes of this picture is the tremendously small thickness of
the arcs. These may be greatly elongated and even form full rings
if the source is larger than the caustic, but their radial size is
really very small. In Figs. \ref{Fig k=2}-\ref{Fig k=3}, we had to
exaggerate the thickness in order to show them in a more evident
way. The next section will be devoted to the calculation of the
length and the thickness of the arcs, i.e. the magnification of
the images.

\subsection{Magnification}

In standard weak field gravitational lensing the magnification is
the ratio between the angular area of an image and the angular
area of the source if no lensing occurred. This does not
necessarily make sense when high deflection takes place, since the
side of the source seen by the black hole is generally different
from the side seen by the observer. Then, if the source does not
emit isotropically, a magnification calculated in the standard way
would not give the correct ratio between the brightness of the
image and that of the source. For example, in the retrolensing
situation, the source is in front of the black hole. So the
photons going toward the black hole leave the source from the side
opposite to the one seen by the observer. For simplicity, in this
section we shall assume that the source emits isotropically. The
formulae can be easily corrected in the case this does not happen.
The isotropic emission hypothesis ensures that the source as seen
by the observer is simply $D_{LS}^2/D_{OS}^2$ smaller than as seen
by an observer in the black hole position in the absence of the
lens.

The angular area of the image in the observer sky is simply
$d\theta_1d\theta_2$. The angular area of the source in the black
hole Boyer-Lindquist coordinates is $|\sin \vartheta_s d\gamma
d\vartheta_s|= d\gamma d\mu_s$, when the source is very far from
the black hole. Then, the magnification matrix is just the
Jacobian matrix of the lens map in the form
\begin{equation}
\left\{
\begin{array}{l}
\gamma=\gamma(\theta_1,\theta_2) \\
\mu_s=\mu_s(\theta_1,\theta_2)
\end{array}
\right. . \label{IdealLens}
\end{equation}

Yet, we have Eqs. (\ref{Lensmu})-(\ref{Lensgamma}) in the form
\begin{equation}
\left\{
\begin{array}{l}
\delta \gamma = \delta \gamma (\delta \psi,\xi) \\
\delta \mu_s = \delta \mu_s (\delta \psi,\xi)
\end{array}
\right.
\end{equation}
and Eqs. (\ref{ImagePos1})-(\ref{ImagePos2}) in the form
\begin{equation}
\left\{
\begin{array}{l}
\theta_1 = \theta_1 (\delta \psi,\xi) \\
\theta_2 = \theta_2 (\delta \psi,\xi)
\end{array}
\right.
\end{equation}

Then we can find the Jacobian matrix of Eq. (\ref{IdealLens}) as
\begin{equation}
\frac{\partial (\gamma,\mu_s)}{\partial (\theta_1,\theta_2)} = a^2
\frac{\partial (\delta \gamma, \delta \mu_s)}{\partial (\delta
\psi,\xi)} \left[ \frac{\partial (\theta_1,\theta_2)} {\partial
(\delta \psi,\xi)} \right]^{-1},
\end{equation}
where we have used the matrix notation
\begin{equation}
\frac{\partial (y_1,y_2)}{\partial (x_1,x_2)}=
\left(
\begin{array}{cc}
\frac{\partial y_1}{\partial x_1} & \frac{\partial y_1}{\partial
x_2} \\
\frac{\partial y_2}{\partial x_1} & \frac{\partial y_2}{\partial
x_2}
\end{array}
\right)
\end{equation}
and we have noted that $d\gamma=a^2 d(\delta \gamma)$ and
$d\mu_s=a^2 d(\delta \mu_s)$.

Then writing the explicit expression of the elements of $J$ is
straightforward, once we correctly take care of all the signs. We
will not write them here, but we shall give the two eigenvalues of
the Jacobian matrix
\begin{eqnarray}
& \lambda_r=& \frac{2D_{OL}}{3\sqrt{3}\epsilon_{cr}} \\
& \lambda_t=& \frac{2D_{OL}}{3\sqrt{3}(1+\epsilon_{cr})} D_0
\end{eqnarray}
with $D_0$ being proportional to the Jacobian studied in the
former section (\ref{Jac2})
\begin{eqnarray}
&&  \!\!\!\!\!\!\! \!\!\!\!\!\!\! D_0 =\pm \sqrt{1-\xi^2} D =
\nonumber \\ && \!\!\!\!\!\!\! \frac{(-1)^k a^2}{9}\left[9\,
\delta\psi- (92-24\sqrt{3}-15k\pi)(1-\xi^2) \right].
\end{eqnarray}

Since $\epsilon_{cr}$ is fixed by the caustic order $k$,
$\lambda_r$ is always positive, while $\lambda_t$ vanishes
whenever $D_0$ does. This condition is fulfilled when Eq.
(\ref{dpsi}) holds, i.e. on critical images. It is possible to
show that the two eigenvectors associated to $\lambda_r$ and
$\lambda_t$ respectively become radial and tangential in the limit
$a\rightarrow 0$. So, when the source is close to a caustic, all
images are elongated in a direction nearly tangential to the
critical curve, as already noticed in the previous subsection. We
shall call $\lambda_r$ and $\lambda_t$ radial and tangential
eigenvalues respectively, though they are such only in the limit
$a\rightarrow 0$, actually.

Finally, we can write the radial and tangential magnification of
the images w.r.t. the source as seen by the observer. These are
\begin{eqnarray}
&& \mu_r=\frac{D_{OS}}{D_{LS}} \frac{1}{\lambda_r}=
\frac{D_{OS}}{D_{LS}} \frac{3\sqrt{3}\epsilon_{cr}}{2D_{OL}}
\label{mur}
\\
&& \mu_t=\frac{D_{OS}}{D_{LS}}
\frac{1}{|\lambda_t|}=\frac{D_{OS}}{D_{LS}}
\frac{3\sqrt{3}(1+\epsilon_{cr})}{2D_{OL} |D_0|}. \label{mut}
\end{eqnarray}

Of course, the total magnification is $\mu=\mu_r\mu_t$. A good
check is to reduce this formula in the Schwarzschild limit to
compare with Refs. \cite{Oha,BCIS,Boz1,BozMan}, and in the
equatorial limit, to compare with Ref. \cite{BozEq}. The first
limit is obtained sending $a$ to zero keeping the source position
$(a^2 \delta \gamma,a^2 \delta \mu_s)$ fixed. This is equivalent
to put $x_c$ to zero in Eqs. (\ref{Eqxi4}), (\ref{Eqpsiimages}).
Then the images are in
\begin{eqnarray}
&\xi=& \pm \delta \gamma/\sqrt{\delta \gamma^2+\delta \mu_s^2} \\
&\delta \psi =& -\frac{16 \delta \mu_s^2}{9(\delta \gamma^2+
\delta \mu_s^2)} \mp \sqrt{\delta \gamma^2 +\delta \mu_s^2}.
\end{eqnarray}
Substituting in Eq. (\ref{mut}), we find
\begin{equation}
\mu_{Sch}= \frac{D_{OS}^2}{D_{LS}^2 D_{OL}^2}\frac{27
\epsilon_{cr}(1+\epsilon_{cr})}{4 a^2 \sqrt{\delta \gamma^2+\delta
\mu_s^2}}.
\end{equation}
Identifying $a^2\sqrt{\delta \gamma^2+\delta \mu_s^2}$ with the
misalignment of the source with the point-like caustic position,
we exactly find the magnification of Refs. \cite{Oha,BozMan}.

The equatorial limit is recovered when $\delta \mu_s=0$. Then we
have two equatorial images plus two non-equatorial images if the
source is inside the caustic. The two equatorial images are
described by
\begin{eqnarray}
&\xi=& \pm 1 \\
&\delta \psi =& -r_c\mp \delta \gamma.
\end{eqnarray}
Inserting these values in Eq. (\ref{mut}), we find
\begin{equation}
\mu_{eq}= \frac{D_{OS}^2}{D_{LS}^2 D_{OL}^2}\frac{27
\epsilon_{cr}(1+\epsilon_{cr})}{4 a^2 |\delta \gamma \pm r_c|},
\end{equation}
which is the leading term close to the equatorial cusps as found
in Ref. \cite{BozEq}.

The magnification of relativistic images is usually very low. This
is expected by the fact that a very small perturbation in the
photon trajectory may completely change its final direction.
Referring to a source at 100AU from Sgr A*, by Eq. (\ref{mur}) we
find a radial magnification
\begin{eqnarray}
&& \mu_r^{(2)}= 5.3\times 10^{-5} \left( \frac{100AU}{D_{LS}}
\right) \label{mur2max}\\
&& \mu_r^{(3)}= 2.3\times 10^{-6} \left( \frac{100AU}{D_{LS}}
\right)
\end{eqnarray}
for the relativistic images of order 2 and 3 respectively. These
very low values, which are independent (at lowest order) of the
distance of the source from the caustic, justify the very thin
arcs of Fig. \ref{Fig Compk23}.

The tangential magnification diverges when the source crosses a
caustic. However, at most the images may merge to form a full
Einstein ring, which gives the maximal tangential magnification.
Dividing the angular area of a circular corona of radius
$3\sqrt{3}(1+\epsilon_{cr})/(2D_{OL})$ and thickness $ \mu_r
(2R_S)/D_{OS}$ by the angular area of the source $\pi
(R_S/D_{OS})^2$, we obtain the maximal total magnification

\begin{eqnarray}
&\mu_{max}^{(2)}=& \frac{1}{ R_S}\frac{D_{OS}^2}{D_{LS}}
\frac{27\epsilon_{cr}(1+\epsilon_{cr})}{D_{OL}^2}= \nonumber
\\
&& 8.7\times 10^{-3} \left( \frac{R_{S}}{R_\odot}
\right)^{-1}\left( \frac{D_{LS}}{100AU} \right)^{-1}.
\label{mut2max}
\\
&\mu_{max}^{(3)}=& \frac{1}{ R_S}\frac{D_{OS}^2}{D_{LS}}
\frac{27\epsilon_{cr}(1+\epsilon_{cr})}{D_{OL}^2}= \nonumber
\\
&& 3.8\times 10^{-4} \left( \frac{R_{S}}{R_\odot}
\right)^{-1}\left( \frac{D_{LS}}{100AU} \right)^{-1}.
\label{mut3max}
\end{eqnarray}
%

\begin{figure}
\resizebox{\hsize}{!}{\includegraphics{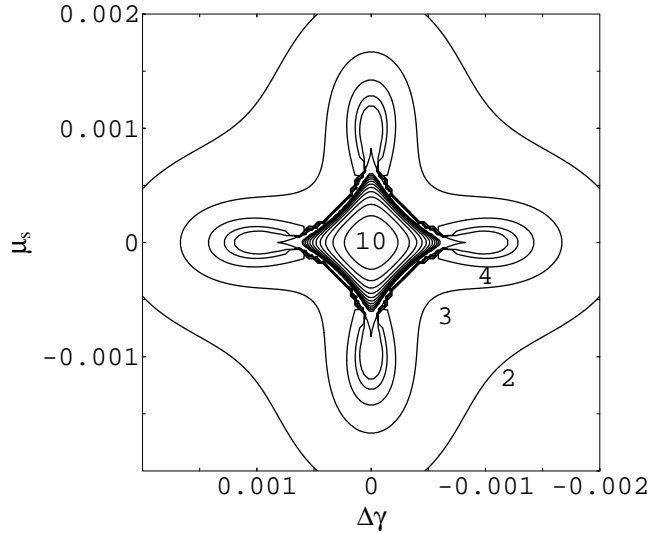}} \caption{
Tangential magnification map centered on the $k=2$ caustic (the
first retro-lensing caustic) for $a=0.02$ and $D_{LS}=100AU$.
 }
 \label{Fig Mag}
\end{figure}

In Fig. \ref{Fig Mag} we show a map of the tangential
magnification centered on the first retro-lensing caustic (the
caustic order is $k=2$) for $a=0.02$. To get the total
magnification map, it is necessary to multiply the values on this
map by the radial magnification (\ref{mur2max}), which is
practically constant in a the neighborhood of the caustic. It is
interesting to note that the tangential magnification is greater
than one in a large region surrounding the caustic. This means
that the cross section for the generation of large relativistic
tangential arcs is quite high. This is a very important
characteristic of relativistic images that we are going to exploit
in the next section when discussing their observability.

For a source at the center of the caustic, we have an Einstein
cross where all images have the same magnification. Then the total
magnification takes a very simple expression
\begin{equation}
\mu_{c}= \frac{D_{OS}^2}{D_{LS}^2D_{OL}^2}
\frac{27\epsilon_{cr}(1+\epsilon_{cr})}{r_c}
\end{equation}
where $r_c$ is the semi-axis of the caustic as defined by Eq.
(\ref{rc}).

\section{Perspectives for observations}

The relativistic images appear just outside the shadow of the
black hole. In order to distinguish them, we need a resolution of
the order of $\mu$as. The present world record has been achieved
with Very Long Baseline Interferometry (VLBI) in the mm band and
amounts to 18 $\mu$as \cite{Krich}. However, in this band, there
are no good compact candidate sources around Sgr A* for
gravitational lensing. More interesting are the infrared and
especially the X-ray band.

\subsection{Infrared band}

In the infrared K-band, centered at $\lambda=2.2$ $\mu$m, the
exinction by interstellar dust allows good observations of the
stellar environment around Sgr A* \cite{Eis}. Many stars have been
detected and followed during their orbital motion around Sgr A*,
providing the best dynamical constraints on its mass distribution.
Surprisingly, these stars are of early spectral types, leaving
open the question on the presence of such young stars in the
galactic center. In the K-band, these stars have magnitudes
between $m_K=13$ and $m_K=16$.

As regards the angular resolution in the K-band, the VLT units can
be combined to perform interferometry observations with an
equivalent baseline of 200 m and a maximal angular resolution of
2.2 mas (http://www.eso.org/projects/vlti). Some space missions
performing nulling interferometry (TPF,
http://www.terrestrial-planet-finder.com; DARWIN,
http://ast.star.rl.ac.uk/darwin) should be launched in the near
future. According to the mission designs, some spacecraft should
fly in formation at distances of the order of tens of meters. A
futuristic development of such idea might lead to much higher
resolutions. The baseline needed for 1$\mu$as resolution is of the
order of hundreds of kilometers. High precision formation flying
may be achieved by laser ranging and microthrusters in the wake of
what is being studied for LISA (http://lisa.nasa.gov), where the
distance between the spacecraft is 5 million km.

Some of the stars around Sgr A* may cross some caustics and
generate bright relativistic images. However, they would be
embedded in the flux coming from Sgr A* environment. In the
quiescent state, Sgr A* flux in the K-band should have $m_K\gtrsim
18.8$ \cite{GheGen}. The infrared emission of Sgr A* is believed
to originate in the inner 10 Schwarzschild radii of the black hole
\cite{YQN}, with the Schwarzschild radius being
\begin{equation}
R_\mathrm{Sch}=\frac{2G M}{c^2} = 1.1 {\times} 10^{10} ~
\mathrm{m}.
\end{equation}
Then it is necessary to establish whether relativistic images can
overcome the background flux. As conservation of surface
brightness holds in gravitational lensing, the relativistic images
must have the same surface brightness of the original source as
seen by the observer,
\begin{equation}
I_S  =  \frac{L_S}{4 \pi D_{OS}^2\Omega_S},
\end{equation}
where $L_S$ the intrinsic luminosity of the source, $D_{OS} \simeq
8$ kpc is the distance to the source and $\Omega_S$ is the angular
area in the sky subtended by the source, which is $\Omega_S\simeq
\pi (R_S)^2/D_{OS}^2$ . For a source of 10 solar radii with
$m_K=15$, the surface brightness of the relativistic images is
four order of magnitudes larger than that of Sgr A*.

However, relativistic images have a tiny angular area and their
contribution to the number of photons collected by a pixel in a
CCD detector may be very small. To get an idea of this fact, let
us consider a CCD detector where every pixel collects energy flux
from an angular area of size $\omega_p^2$. If a relativistic
image, in the form of a tangential arc of angular thickness
$\omega_{arc} = \mu_r (2R_S)/D_{OS}$, lies on the pixel area, the
flux received is $S \propto  I_S (\omega_p {\times}\omega_{arc})$.
On the other hand, the noise coming from the environment of Sgr A*
to the single pixel is $ N \propto  I_{Sgr} \omega_p^2$. Then, the
signal-to-noise ratio for a single pixel is
\begin{eqnarray}
\frac{S}{N} & = &\frac{I_S}{I_{Sgr}} \frac{\omega_p}{\omega_{arc}
} \label{inf1} \\ & \simeq & 2.2 \left( \frac{\omega_p}{1\mu
as}\right)^{-1} \left( \frac{R_S}{10R_\odot}\right)^{-1}\left(
\frac{D_{LS}}{100AU}\right)^{-1}.
\end{eqnarray}
i.e. a pixel with a tangential arc receives only twice more K-band
photons than other pixels. The $S/N$ could be improved by taking a
smaller pixel size or stars with lower radii and higher
brightness, but we cannot go very far. Taking into account
absorption by the matter surrounding the black hole that would
surely take place and fluctuations in the surface brightness of
Sgr A*, we doubt that relativistic images of stellar sources can
be actually detected in the K-band by present or near future
technology.

\subsection{X-ray band}

In the X-rays, Chandra is leading very important observations
discovering the physics of high energy electromagnetic sources in
the central region of the Galaxy \cite{Bagan} with a resolution of
the order of 0.5 arcseconds. The space mission project MAXIM
(http://maxim.gsfc.nasa.gov) will represent a major leap toward
high resolution, reaching the striking resolution of 0.1 $\mu$as.
With such observational facility, a complete and detailed imaging
of the black hole will be possible.

It is very interesting to consider that Sgr A* luminosity in the
2-10 keV band is $2 {\times} 10^{33}$ ergs s$^{-1}$, which is much
lower than expected if the black hole were accreting at the
Eddington rate, i.e $L = 3{\times} 10^{44}$ ergs s$^{-1}$
\cite{Bagan}. There are several models for Sgr A* accretion.
Models based on Bondi accretion (spherically symmetric inflow)
predict the X-ray emission to be created in a region of the order
of $10^2 R_{Sch}$ \cite{Melia}. Models based on advection
dominated flow \cite{ADAF} predict the emission to be dominated by
cooler gas at larger radii, of the order of $10^4 R_{Sch}$.

Indeed, many X-ray sources have been detected in the neighbourhood
of Sgr A*, with a luminosity comparable or even slightly higher
than the supermassive black hole \cite{Muno}. These sources are
probably Low Mass X-ray Binaries (LMXB) which seem very numerous
in the galactic center. The situation seems really appealing,
since we have a population of bright compact sources, with
possibly poor contamination from the intrinsic luminosity of Sgr
A*. It is believed that the most of the X-ray emission from a LMXB
comes from a region of tens of kilometers. Then we can assume $R_S
\simeq 10^2~\mathrm{km}$, with an X-ray flux of $ L_S\simeq
L_{Sgr}= 2{\times} 10^{33}$ ergs s$^{-1}$. If we consider an
emitting region of $10^2 R_{Sch}$ for Sgr A*, there are 14 orders
of magnitude between the surface brightness of an LMXB and the
surface brightness of the X-ray emission of the supermassive black
hole environment. Then, the signal-to-noise ratio for a single
pixel imaging a tangential arc, Eq.~(\ref{inf1}), is
\begin{equation}
\frac{S}{N} = 0.9 {\times} 10^{6} \left( \frac{\omega_p}{1\mu as}
\right)^{-1} \left( \frac{R_S}{100km}\right)^{-1} \left(
\frac{D_{LS} }{100AU}\right)^{-1}
\end{equation}
i.e. the signal in a pixel touched by a tangential arc is nearly 6
order of magnitudes higher than the noise from Sgr A* environment
for a detector with the accuracy of $1~\mu$as. The contamination
from Sgr A* environment seems to be completely under control.
Relativistic gravitational self-lensing of Sgr A* would just
produce relativistic Einstein rings with the same surface
brightness of Sgr A*. The only serious danger for photons coming
from outside Sgr A* and deflected by the central black hole is
absorption by the matter surrounding it. However, even if some
absorption certainly occurs, it seems difficult to fill a gap of
so many orders of magnitudes between the surface brightness of the
relativistic images and that of Sgr A* without affecting the
luminosity of Sgr A* as well.

Of course, the idea firstly proposed by Rauch \& Blandford that
some X-ray flares may be explained by lensing of nearby sources is
fully plausible in this scenario \cite{RauBla}.

\section{Conclusions}

In this paper we have made an analytical treatment of
gravitational lensing by Kerr black holes in the strong deflection
limit. In order to achieve our objective we have made three
approximations.

The first one is the strong deflection limit approximation for all
radial integrals. This is just an expansion of the elliptic
integrals that result from integrations over the full radial
motion of a photon. Restricting to photons suffering a very large
deflection, for all the radial integrals we have only kept the
leading term diverging as $\log \epsilon$ and the constant term
($\epsilon$ being the separation between the image and the shadow
border as seen by the observer). As shown in several articles,
this limit gives a very good approximation starting from photons
deflected by an angle of order $\pi$ \cite{Dar,Oha,BCIS,S1-S14}.

The second approximation has been to consider only small values of
the black hole spin $a$. This has allowed us to take the
Schwarzschild gravitational lensing as a starting point for the
derivation of the corrections due to the presence of an intrinsic
angular momentum of the black hole. As far as we could compare our
results with available exact ones, we have verified a considerably
wide applicability range of our approximation. For the first
relativistic images, we can safely apply our treatment up to
$a=0.1$ ($a=0.5$ being the extremal Kerr black hole in our units).

The third restriction concerns the position of source and
observer. Besides the limitation to far sources and observers
($D_{LS},D_{OL} \gg R_{Sch}$), we have also restricted to
equatorial observers. This considerably simplifies all
calculations without affecting the complete investigation of the
most significant physical situation, namely the black hole in Sgr
A*. In fact, it is natural to assume that the equatorial plane of
this supermassive black hole coincides with the galactic plane. In
any case, a full investigation stepping beyond these restrictions
is in progress.

The first achievement of this paper has been the analytical
description of the Kerr caustics. At the first order in $a$ we
find that they are just shifted along the equatorial plane still
remaining point-like, while at second order they are resolved into
typical diamond-shaped figures. We are thus able to calculate the
position and the extension of the caustics for any order of
relativistic images (as long as we remain in the perturbative
regime). As stated in Sect. IV, the strong deflection limit
treatment does not cover the primary caustic ($k=1$ in our
formulae), since this caustic is formed in the weak field regime
for $D_{LS} \gg 1$ \cite{RauBla,Ser}. With no regard to the
lensing regime, the effect of the angular momentum of the
deflector is similar, with caustics getting a diamond-shaped
structure and drifting from the optical axis. Whereas
magnification effects due to the primary caustic are very large,
its size is very small, so that the creation of additional pairs
of images (which is the most evident manifestation of the presence
of a non-negligible spin) is very difficult to achieve. The
significant extension of relativistic caustics strongly enhances
the cross-section for additional images and puts them in a much
better position for testing the Kerr nature of the black hole. It
has to be remarked that our perturbative investigation still
leaves open the possibility that metamorphosis may occur at large
values of $a$. Though this has been numerically excluded for the
primary caustic \cite{RauBla}, it is possible that higher order
caustics develop more complicated structures in a strongly
non-pertubative regime.

The second achievement has been the analytical inversion of the
lens mapping near the caustics, which has allowed us to draw
fascinating pictures of the relativistic images generated by a
hypothetical source close to a relativistic caustic.

However, the most important result has been the possibility of
doing concrete analytical estimates of the size and luminosity of
the relativistic images. The LMXBs surrounding Sgr A* provide an
ideal population of sources, which may eventually bump into a
relativistic caustic and generate appreciable relativistic images.
This is because they are compact sources with very high surface
brightness in the X-rays, compared to that of Sgr A*. This seems
not to be the case for stellar sources in the infrared K-band,
which have a too small surface brightness. Using our formulae for
the magnification of relativistic images, we are entitled to claim
that future space missions performing X-ray interferometry with
resolutions of the order of 1 $\mu$as will see these relativistic
images with high probability.

\appendix

\section{Resolution of radial integrals}

In this appendix we recall the SDL technique used in Ref.
\cite{Boz1} to solve radial integrals, applying it to the
integrals that appear in the geodesics equations. We rewrite them
here for an easier reading
\begin{eqnarray}
&& I_1=2\int\limits_{x_0}^{\infty} \frac{dx}{\sqrt{R}}
\label{I2App} \label{I1App}\\
&& I_2=2\int\limits_{x_0}^{\infty} \frac{x^{2}+a^{2}-a J}{\Delta
\sqrt{R}} dx.
\end{eqnarray}

First we change the integration variable from $x$ to $z$ by the
transformation
\begin{equation}
x=\frac{x_0}{1-z}. \label{xtoz}
\end{equation}
As a consequence, the integration domain $[x_0,\infty]$ becomes
$[0,1]$.

Then, each of the integrals $I_1$, $I_2$ can be written in the
form
\begin{eqnarray}
&& I_i=\int\limits_0^1 R_i(z)f(z) dz \\
&& f(z)=\frac{1}{\sqrt{R(z)}},
\end{eqnarray}
where  the two functions $R_i(z)$ can be easily read by Eqs.
(\ref{I1})-(\ref{I2}) taking into account the Jacobian of the
transformation (\ref{xtoz}):
\begin{eqnarray}
& R_1(z)=&\frac{2x_0}{(1-z)^2}
\\
& R_2(z)=& \frac{2x_0}{(1-z)^2}\frac{x_0^{2}+(1-z)^2(a^{2}-a J)}{
x_0^2-x_0(1-z)+a^2(1-z)^2}.
\end{eqnarray}

Now we consider the expansion of $R(z)$ in a neighborhood of
$z=0$. Since $z=0$ means $x=x_0$ and $x_0$ is a root of $R(x)$, we
deduce that $R(z=0)=0$. Then the expansion of $R(z)$ reads
\begin{equation}
R(z)=\alpha z + \beta z^2 +o(z^2),
\end{equation}
where the coefficients of the expansion are
\begin{eqnarray}
&\alpha=& x_0\left[(a-J)^2+Q+2(a^2-J^2-Q)x_0\right. \nonumber \\
&& \left. +4x_0^3 \right] \\
&\beta=& x_0\left[(a-J)^2+Q+3(a^2-J^2-Q)x_0\right. \nonumber \\
&& \left. +10x_0^3 \right]
\end{eqnarray}

We use this expansion to define
\begin{equation}
f_0(z)=\frac{1}{\sqrt{\alpha z +\beta z^2}}.
\end{equation}

The radial integrals can be split in two pieces
\begin{eqnarray}
&& I_i=I_{i,D}+I_{i,R} \\
&& I_{i,D}=\int\limits_0^1 R_i(0)f_0(z)dz \\
&& I_{i,R}=\int\limits_0^1 \left[ R_i(z)f(z)-R_i(0)f_0(z) \right]
dz.
\end{eqnarray}

The first integral gives the result
\begin{equation}
I_{i,D}=\frac{2R_i(0)}{\sqrt{\beta}}\log
\frac{\sqrt{\beta}+\sqrt{\alpha+\beta}}{\sqrt{\alpha}}.
\end{equation}
$\alpha$, $\beta$ and $R_i(0)$ are known functions of $x_0$, $J$,
$Q$ and $a$. Now we can use the SDL parameterizations
(\ref{JSFL}), (\ref{QSFL}), (\ref{x0SFL}) for all these
quantities, so that they become functions of $\xi$, $\delta$ (or
equivalently $\epsilon$) and $a$. Then, in the spirit of SDL
approximation, we keep the leading order in $\delta$, which goes
as $\log \delta$, and the next-to-leading order which is constant
in $\delta$. Finally, we expand the obtained expression to second
order in $a$.

As regards the integrals $I_{i,R}$, the integrand function is
regular in the whole integration domain. Sending $\delta$ to zero,
the integrand does not diverge. So, this integral contributes to
the SDL expansion with another constant in $\delta$ plus higher
order terms that we can neglect. It is convenient to make the
second order expansion in $a$ before the integration, in order to
have a sum of easily integrable functions. We can then add the
result of the integral $I_{i,R}$ to the integral $I_{i,D}$, to
reconstruct the full SDL formulae for radial integrals:
\begin{eqnarray}
& I_1=&- a_1 \log \delta+ b_1 \\
& I_2=&- a_2 \log \delta+ b_2
\end{eqnarray}

The coefficients expanded to second order in $a$ read
\begin{eqnarray}
& a_{1}=&\frac{4}{3 \sqrt{3}}+\frac{16}{27} a \xi+\frac{8}{27
\sqrt{3}} (1+3\xi^2) a^2 \\%
& b_1=&a_1 \log \left[6\left( \sqrt{3}-1 \right)^2\right]
\nonumber \\ &&- \frac{8}{27\sqrt{3}}a^2 \left( 5-2\sqrt{3}
\right) (1-\xi^2)
\end{eqnarray}
\begin{eqnarray}
& a_{2}=&\frac{4}{\sqrt{3}}+\frac{8}{3}a \xi+ \frac{8
\sqrt{3}}{27} (1+7 \xi^2) a^2 \\%
& b_2  =& [a_2+4(1+2a^2)] \log\left[ 6({\sqrt 3}-1)^2\right]
\nonumber \\ && -2(1+2a^2) \log48 -\frac{8}{3\sqrt{3}}a\xi
(3\sqrt{3} - 2 )\nonumber \\ && +\frac{8}{27}
a^2[19\sqrt{3}-12+\xi^2(14 -25\sqrt{3})].
\end{eqnarray}

The separation of $I_i$ into $I_{i,D}$ and $I_{i,R}$ is necessary
to isolate the term generating the $\log \delta$ into an easier
integral.

\section{Angular integrals}

This appendix is devoted to the resolution of the angular
integrals
\begin{eqnarray}
&& J_1=\pm \int \frac{1}{\sqrt{\Theta}} d \vartheta \\
&& J_2=\pm \int \frac{csc^2\vartheta}{\sqrt{\Theta}} d \vartheta .
\end{eqnarray}
Introducing the variable $\mu=\cos \vartheta$, the two integrals
become
\begin{eqnarray}
&& J_1=\pm \int \frac{1}{\sqrt{\Theta_\mu}} d \mu \label{J1mu}\\
&& J_2=\pm \int \frac{1}{(1-\mu^2)\sqrt{\Theta_\mu}} d \mu
,\label{J2mu}
\end{eqnarray}
where
\begin{eqnarray}
& \Theta_\mu=&a^2(\mu_-^2+\mu^2)(\mu_+^2-\mu^2) \\
& \mu_\pm^2=&\frac{\sqrt{b_{JQ}^2+4a^2Q_m}\pm b_{JQ}}{2 a^2} \\
& b_{JQ}=& a^2-J_m^2-Q_m,
\end{eqnarray}
and we have already replaced $J$ and $Q$ with $J_m$ and $Q_m$,
coherently with the fact that we only retain terms which are
logarithmically diverging or constant in $\delta$ (or equivalently
$\epsilon$).

$\Theta_\mu$ has two zeros in $\mu=\pm \mu_+$. Then the photon
performs symmetric oscillations of amplitude $\mu_+$ w.r.t. the
equatorial plane. It is useful to write the explicit expressions
of $\mu_+$ and $\mu_-$ in terms of $a$ and $\xi$, using Eqs.
(\ref{Jma2})-(\ref{Qma2}) and expanding to second order in $a$. We
find
\begin{eqnarray}
& \mu_{+}=&\sqrt{1-\xi^2}\left(1+\frac{4 a \xi }{3
\sqrt{3}}-\frac{8 a^2 \xi^2 }{27} \right) \\
& \mu_{-}=&\frac{3 \sqrt{3}}{2 a}-2 \xi +\frac{a(4-8 \xi^2)}{3
\sqrt{3}} \nonumber \\ &&+\frac{4 a^2 \xi (12-17 \xi^2)}{27}.
\end{eqnarray}

The oscillation amplitude is $\sqrt{1-\xi^2}$ plus corrections due
to the black hole spin. This is coherent with the fact that for a
photon reaching the observer from the equatorial plane ($\xi=\pm
1$) the amplitude of the oscillation goes to zero. On the other
hand, a photon moving on a polar orbit ($\xi=0$) performs
oscillations of maximal amplitude, touching the poles of the black
hole. Now, to perform the angular integrals, it is wise to expand
the integrands to second order in $a$ and then integrate. Then,
the primitive functions read
\begin{eqnarray}
& F_{J_1}(\mu)& = \frac{2}{3\sqrt{3}}\arcsin \frac{\mu}{\sqrt{1-\xi^2}} \nonumber \\
&& +\frac{8}{27} a\xi \left[\arcsin
\frac{\mu}{\sqrt{1-\xi^2}}-\frac{\mu}{\sqrt{1-\mu^2-\xi^2}}
\right] \nonumber \\
&& +\frac{2a^2}{81\sqrt{3}} \left[ (33\xi^2-9) \arcsin
\frac{\mu}{\sqrt{1-\xi^2}} \right. \nonumber \\
&& \left.
+\frac{\mu(\mu^4+2\mu^2(\xi^2-1)+1+6\xi^2-7\xi^4)}{(1-\mu^2-\xi^2)^{3/2}}
\right]
\end{eqnarray}
\begin{eqnarray}
& F_{J_2}(\mu)& = \frac{2}{3\sqrt{3}\xi}\arctan \frac{\mu
\xi}{\sqrt{1-\mu^2-\xi^2}} \nonumber \\
&& +\frac{8}{27\xi^2} a \left[\arctan \frac{\mu
\xi}{\sqrt{1-\mu^2-\xi^2}}-\frac{\mu \xi}{\sqrt{1-\mu^2-\xi^2}}
\right] \nonumber \\
&& +\frac{4a^2}{81\sqrt{3}} \left[
\frac{12-21\xi^2+20\xi^4}{\xi^3} \arctan \frac{\mu
\xi}{\sqrt{1-\mu^2-\xi^2}} \right. \nonumber \\
&& \left.
+\frac{4\mu(1-\xi^2)(3\mu^2-3+4\xi^2)}{\xi^2(1-\mu^2-\xi^2)^{3/2}}
\right. \nonumber \\ && \left. + \arctan \frac{\mu
}{\sqrt{1-\mu^2-\xi^2}} \right]
\end{eqnarray}

The integration limits are the values of $\mu$ at the observer and
source position. The observer is at $\mu_o=0$, since it lies on
the equatorial plane, while the source is in $\mu_s=\cos
\vartheta_s$. Notice that the choice of an equatorial observer
leads to a considerable simplification, since
$F_{J_1}(0)=F_{J_2}(0)=0$. Moreover, we have to consider that
during the photon motion, $\mu$ may perform several oscillations
between $-\mu_+$ and $\mu_+$, depending on how many loops the
photon makes around the black hole before escaping. So, we have to
add an arbitrary integer number $m$ of integrals covering the
whole domain $[-\mu_+,\mu_+]$. The final results read
\begin{eqnarray}
& J_1=&\pm F_{J_1}(\mu_s) \nonumber \\
&& +\frac{2m\pi}{3\sqrt{3}}\left[1+\frac{4a\xi}{3\sqrt{3}} +\frac{a^2(11\xi^2-3)}{9} \right] \\
& J_2=&\pm F_{J_2}(\mu_s)+\frac{2m\pi}{3\sqrt{3}\xi} \left[ 1+ \frac{4a}{3\sqrt{3}\xi}  \right. \nonumber \\
&& \left. + \frac{2a^2(12-21\xi^2+\xi^3+20\xi^4)}{27\xi^2}
\right],
\end{eqnarray}
where the minus signs hold if the photon initially increases its
latitude and the plus signs hold if the latitude decreases
initially.

\begin{acknowledgments}
V.B. and M.S. thank all the participants of the workshop
"Gravitational Lensing in the Kerr Spacetime Geometry", held at
the American Institute of Mathematics in Palo Alto (CA), for
invaluable discussions and precious interaction, and in particular
Simonetta Frittelli and Arlie Petters for the kind invitation.

V.B. and G.S. acknowledge support for this work by MIUR through
PRIN 2004 ``Astroparticle Physics'' and by research fund of the
Salerno University. F.D.L.'s work was performed under the auspices
of the EU, which has provided financial support to the ``Dottorato
i Ricerca Internazionale in Fisica della Gravitazione ed
Astrofisica'' of the Salerno University, through ``Fondo Sociale
Europeo, Misura III.4''.
\end{acknowledgments}

\end{document}